\documentclass{article}
\usepackage{cancel}
\usepackage{amssymb,amsmath}
\usepackage{wrapfig}
\usepackage{mathrsfs}
\usepackage{amsbsy}
\usepackage{color}
\usepackage{graphicx}
\usepackage[colorlinks=true, pdfstartview=FitV, linkcolor=blue, citecolor=blue, urlcolor=blue]{hyperref}

\usepackage{verbatim}
\newcommand\DD {{\cal D}}
\def\wt{\widetilde}

\def\wh{\widehat}

\def\ds{\displaystyle}
\def\supp{\mathrm {supp}}

\renewcommand{\theequation}{\arabic{section}-\arabic{equation}}
\makeatletter
\@addtoreset{equation}{section}
\makeatother
\def\tr{\mathrm {Tr}}
\def\le{\left}
\def\ri{\right}
\def\QED{{\bf Q.E.D.}\par\vskip 5pt}

\def\gg{\mathfrak g}
\def\bc{\begin{corollary}}
\def\ec{\end{corollary}}
\def\&{&{\hskip -20pt}}

\def\ov{\overline}
\def\br{\begin{remark}\rm\small}
\def\1{{\bf 1}}
\def\er{\end{remark}}
\def\bt{\begin{theorem}}
\def\et{\end{theorem}}

\def\bx{\begin{example}}
\def\ex{\end{example}}
\def\bd{\begin{definition}}
\def\ed{\end{definition}}
\def\bp{\begin{proposition}\rm}
\def\bl{\begin{lemma}\em}
\def\el{\end{lemma}}
\def\ep{\end{proposition}}
\def\bea{\begin{eqnarray}}
\def\eea{\end{eqnarray}}
\def \pa{\partial}
\def\C{{\mathbb C}}
\def\R{{\mathbb R}}
\def\N{{\mathbb N}}
\def\Z{{\mathbb Z}}

\newtheorem{problem}{Problem}[section]
\newtheorem{theorem}{Theorem}[section]
\newtheorem{example}{Example}[section]
\newtheorem{coroll}{Corollary}[section]
\newtheorem{examps}{Examples}[section]

\newtheorem{assumption}{Assumption}
\newtheorem{lemma}{Lemma}[section]
\newtheorem{remark}{Remark}[section]
\newtheorem{remarks}[remark]{Remarks}
\newtheorem{proposition}{Proposition}[section] 
\newtheorem{definition}{Definition}[section]

\def\br{\begin{remark}}
\def\er{\end{remark}}
\def\bt{\begin{theorem}}
\def\et{\end{theorem}}
\def\bc{\begin{coroll}}
\def\ec{\end{coroll}}
\def\brs{\begin{remarks} \rm\
\begin{enumerate}}
\def\ers{\end{enumerate}\end{remarks}}
\def\bl{\begin{lemma}}
\def\el{\end{lemma}}
\def\bxs{\begin{examps}. \rm\begin{enumerate}}
\def\exs{\end{enumerate}\end{examps}}
\def\bd{\begin{definition}}
\def\ed{\end{definition}}
\def\bp{\begin{proposition}}
\def\ep{\end{proposition}}
\def\be{\begin{equation}}
\def\ee{\end{equation}}

\def\d{{\rm d}}
\def\bea{\begin{eqnarray}}
\def\eea{\end{eqnarray}}
\def\beas{\begin{eqnarray*}}
\def\eeas{\end{eqnarray*}}
\def \pa{\partial}
\def\iint{\int\!\!\!\!\int}
\def\C{{\mathbb C}}
\def\L{\mathcal L}
\def\R{{\mathbb R}}
\def\N{{\mathbb N}}

\def\Z{{\mathbb Z}}
\def\ZZ{{\boldsymbol \Gamma}}

\date{}
\begin{document}
\baselineskip 16pt plus 1pt minus 1pt
\vspace{0.2cm}
\begin{center}
\begin{Large}
\textbf{Strong asymptotics for Cauchy biorthogonal polynomials  with application to the Cauchy two--matrix model}
\end{Large}\\
\bigskip
\begin{large} {M.
Bertola$^{\dagger}$,  M. Gekhtman$^{\ddagger}$, J. Szmigielski }$^{\star}$
\end{large}
\\
\bigskip
\begin{small}
$^{\dagger}$ {\em CRM,
Universit\'e de Montr\'eal and Department of Mathematics and
Statistics, Concordia University}\end{small}

\begin{small}
$^{\ddagger}$ {\em Department of Mathematics, University of Notre Dame}\end{small}

\begin{small}
$^{\star}$ {\em Department of Mathematics and Statistics, University of Saskatchewan}\end{small}
\end{center}
\bigskip

\begin{center}{\bf Abstract}\end{center}
We apply the nonlinear steepest descent method to a class of $3\times 3$ Riemann-Hilbert problems introduced in connection with the Cauchy two-matrix random model.  The general case of two equilibrium  measures supported on an arbitrary number of intervals is considered. In this case, we solve the Riemann-Hilbert problem for the outer parametrix in terms of sections of a spinorial line bundle 
on a three--sheeted Riemann surface of arbitrary genus and establish strong asymptotic results for the Cauchy biorthogonal polynomials. 

\tableofcontents

\section{Introduction}
In this paper, we study asymptotic behavior of a class of $3\times 3$ Riemann-Hilbert problems motivated by
the recently introduced two-matrix random model \cite{Bertola:CauchyMM}. 
The model consists of two random Hermitean positive-definite matrices $M_1, M_2$ of size $n\times n$ with the probability measure
\be
\label{2matrix}
\d \mu(M_1,M_2)= \frac 1{\mathcal Z_n} \frac {\d M_1 \d M_2}{\det(M_1+M_2)^n} {\rm{e}}^{-N\tr(U(M_1))}{\rm{e}}^{-N\tr(V(M_2))}
 \ee
 where $U,V$ are scalar functions defined on $\R_+$.
The model was termed the {\em Cauchy matrix model} because of the shape of the coupling term. Similarly to the case of the Hermitean one-matrix models for which the spectral statistics is expressible in terms of appropriate biorthogonal polynomials \cite{Mehtabook}, this two-matrix model is solvable with the help of a new family of biorthogonal polynomials named {\em the Cauchy biorthogonal polynomials} \cite{Bertola:CBOPs}.

The Cauchy biorthogonal polynomials are  two sequences of monic polynomials $(p_j(x))_{j=0}^\infty, (q_j(y))_{j=0}^\infty$ with deg $p_j=$deg $q_j=j$ that satisfy
\be
\iint_{\R_+\times \R_+}p_j(x)q_k(y) \frac {{\rm e}^{-N(U(x)+V(y))}}{x+y} \d\, x \d\, y= h_k\delta_{jk}\ ,\ \ \forall j,k \ge 0\ ,h_k>0\ .
\ee
These polynomials were studied in \cite{Bertola:Cubic} in relation with the spectral theory of the cubic string.

In yet another application, in analogy to the  moment problem approach to the Camassa-Holm peakons \cite{bss-moment, bss-stieltjes}, the predecessors of
the Cauchy biorthogonal polynomials were used to study the peakon solutions of the  Degasperis-Procesi wave equation \cite{ls-cubicstring, ls-invprob}. More generally,
the Cauchy biorthogonal polynomials are expected to play a role in a variety of inverse problems for the third order differential operators \cite{keivan}.

The main features of Cauchy BOPs can summarized as follows:
\begin{enumerate}
\item they solve a four-term recurrence relation;
\item their zeroes are positive and simple;
\item their zeroes have the interlacing property;
\item they  satisfy Christoffel--Darboux identities;
\item they can be characterized by a pair of $3\times 3$ Riemann--Hilbert problems;
\item the solution to the corresponding Riemann--Hilbert problem yields all kernels for the correlation functions of the Cauchy two-matrix model;
\end{enumerate}

Items 1-5 above have been addressed in \cite{Bertola:CBOPs} while item 6 was explained in \cite{Bertola:CauchyMM}. In the present paper, we apply the Deift-Zhou steepest descent method to the asymptotic analysis of the Riemann-Hilbert problem with the view towards applications to the biorthogonal polynomials and the spectral statistics of  the Cauchy  two-matrix model.

The paper is organized as follows. In Section \ref{CBOPs}, we 
set up a Riemann--Hilbert problem that characterizes the biorthogonal polynomials and that is essential in evaluating the correlation kernels for the associated matrix model.

In Section \ref{seclargen},  we recall the results of \cite{BertolaBalogh, Bertola:CauchyMM} where a relevant potential theory problem was set up and solved. The resulting equilibrium measures are the key ingredient in the construction of the $\gg$--functions that  pave the way for  the Deift--Zhou steepest descent method.

The central section of the paper is Section \ref{DZdescent}, which deals with the nonlinear steepest descent analysis. We consider the general case in which the two equilibrium  measures are supported on an arbitrary number of intervals. This prompts the use  of a higher genus three--sheeted Riemann surface.  
Note that a Riemann surface of a similar structure was recently used in \cite{Suetin} to study Hermite-Pade approzimations of pairs of functions that form generalized Nikishin systems.

We solve the Riemann-Hilbert problem for the outer parametrix in terms of sections of a spinorial line bundle in the spirit of \cite{BertolaAdvances, Bertola:EffectiveIMRN}). Much of the effort goes towards showing that the model problem for the outer parametrix always admits a solution (Proposition \ref{taufunc}, Theorem \ref{nonzerotau}). 

 It is perhaps worth mentioning that in \cite{MoKuDu} (Section 8) the authors approached a similar problem of solving a $4\times 4$ RHP for multiple orthogonal polynomials that arise in the analysis of the two-matrix models with interaction ${\rm e}^{-N\tr (M_1 M_2)}$ instead of the Cauchy interaction in \eqref{2matrix}. Though similar,  
their approach differs in that they use the theorem on existence of meromorphic differentials on a Riemann surface without providing explicit formul\ae\ in terms of Theta functions, contrary to the present work.  Also we work with arbitrary two measures without being restricted to the choice of a quartic potential.   
Section \ref{universality} uses the asymptotic analysis in the previous sections  to discuss universality results for the spectral statistics of the two-matrix model. We find that individual spectral statistics exhibits the same universality phenomena as in the one-matrix model. We expect that 
 the Cauchy two-matrix model might produce new universality classes in the case when supports of both equilibrium contain the origin. This case, however, is not considered in the present paper due to assumptions on the potentials (Assumption \ref{ass1}). Relaxing this assumption requires a generalization of the potential theory problem studied in \cite{BertolaBalogh}.

The Appendices contain a discussion of a specific genus zero example (Appendix \ref{DoubleJac}) as well as  notations and essential information regarding Theta functions (Appendix \ref{se:notation}).

{\em Acknowledgements.}  This paper was completed at the Banff International Research Station. We thank BIRS for the hospitality and for providing excellent work conditions. We also thank Dima Korotkin for helpful discussions concerning spin bundles. M. B. and J.S. acknowledge a support by Natural Sciences and Engineering Research Council of Canada. M.G. is partially supported by the NSF grant DMS-1101462.

\section{Riemann-Hilbert problem for Cauchy biorthogonal polynomials}
\label{CBOPs}

In the case of ordinary orthogonal polynomials, a well-known characterization in terms of a Riemann-Hilbert problem
was obtained in \cite{FIK0}. Here we present a similar characterization following \cite{Bertola:CBOPs}.

For symmetry reasons, we define 
\be
V_{_1}(z):= U(z)\ ,\qquad V_{_2}(z):= V(-z).
\ee
Following \cite{BertolaBalogh}
we consider potentials $V_j$ that are subject to
\begin{assumption} The potentials $V_j(z)$ satisfy:
\label{ass1}
\begin{itemize}
\item $V_j(z)$ is a {\bf real analytic function} on $(-)^{j+1}\R_+$, $j=1,2$,
\item the growth-conditions
\be
V_j(x)=-a_j \ln |x|+\mathcal{O}(1) \, \quad \text{as  } x\rightarrow 0, \quad a_j>1,\qquad \lim_{x\to(-1)^{j+1}\infty} \frac {V_j(x)}{\ln|x|} = +\infty\label{growthcondV},
\ee
\item the derivatives $V_j'(z)$ are meromorphic on a strip containing the whole real axis.
\end{itemize}
\end{assumption}
\bx
The typical examples are potentials of the form 
\be
V_j(x)= -a_j \ln |x| + P_j(x)\ ,\ \ a_j>1
\ee
where $P_j$ are real polynomials such that $\lim_{x\to (-)^{j+1}\infty} P_j(x) =+\infty $. 
\ex

The relevant Riemann--Hilbert problem that characterizes CBOPs is 
\begin{problem}
\label{RHPp}
Find a matrix $\Gamma(z)$ such that 
\begin{enumerate}
\item $\Gamma(z)$ is analytic in 
$\C \setminus \R$, 
\item 
$\Gamma(z)$ satisfies the jump conditions 
\bea 
\label{eq:RHphat}
\Gamma(z) _+ &\&=  \Gamma(z)_- \le[
\begin{array}{ccc}
1 & {\rm e}^{-N  V_1(z)}   & 0\\
0&1&0\\
0&0&1
\end{array}
\ri]
\ ,\ \ z\in \R_+\\
 \Gamma(z) _+ &\&=  \Gamma(z)_- \le[
\begin{array}{ccc}
1 &  0 & 0\\
0&1& {\rm e}^{-N V_2(z)} \\
0&0&1
\end{array}
\ri]
\ ,\ \ z\in  \R_-, 
\eea
where the negative axis is oriented towards $-\infty$, 

\item at $z=\infty$ 
\bea \label{eq:Gammahat-as}
 \Gamma(z) =  \le(\1  + \mathcal O\le(\frac 1 z\ri)\ri)\le[
\begin{array}{ccc}
z^n & 0&0 \cr
0 &1 &0\cr
0 &0& \frac 1{ z^{n}} 
\end{array} 
\ri]\ ;
\eea
\item near $z=0$ 
\bea
\qquad \Gamma(z) = \le[\mathcal O(1), \mathcal O(\ln |z|), \mathcal O(\ln ^2|z|)\ri]. \label{growth}
\eea  
\end{enumerate}

\end{problem}
\br
\label{strongg}
The growth condition at $z=0$ in eqs. (\ref{growth}) can be replaced by $\mathcal O(1)$ if the densities ${\rm e}^{- N V_j}$ vanish at $x=0$. In fact, after Assumption \ref{ass1} is put in place, we have $ {\rm e}^{- N V_j(x)}= \mathcal O\le(|x|^{a_j N }\ri)$ and this (using Plemelj formul\ae\ for the local model of $\Gamma$ at $z=0$) implies $\Gamma(0) = \mathcal O(1)$.
\er

In \cite{Bertola:CBOPs} (Section 6.2) it was shown that  (adapting to the present notation
\footnote{\label{foot6}In loc. cit. the matrix that we denote here by $\Gamma$ corresponds to $\wh \Gamma$; the precise relation is  
\be
\Gamma(z) = {\rm diag} (1, -2\pi i, -(2\pi i)^2)  \wh \Gamma(z) {\rm diag} \le(1, -\frac 1{2\pi i}, -\frac 1{(2\pi i)^2}\ri)
\ee})
\bea
&\& \Gamma(z) =\label{gammamain} \\
&\& \begin{bmatrix}
p_n(z) & \ds \frac 1 {2i\pi} \int_{\R_+}\!\!\!\! \frac {p_n(x) {\rm e}^{-NV_{_1}(x)}\d x}{x-z} & \ds \frac 1 {(2i\pi)^2} \int_{\R_+}\int_{\R_-} \frac {p_n(x) {\rm e}^{-NV_{_1}(x) -NV_{_2}(y)}\d x\d y}{(y-z)(x-y)} \\
2i\pi \wh p_{n-1}(z) & 1\!+\!\ds  \int_{\R_+} \!\!\!\! \frac{\wh p_{n-1}(x) {\rm e}^{-NV_{_1}(x)}\d x}{x-z} 
& \ds  \int_{\R_+}\!\!\int_{\R_-}\!\!\!\!\!  \frac {\wh p_{n-1}(x) {\rm e}^{-NV_{_1}(x) -NV_{_2}(y)}\d x\d y}{2i\pi (y-z)(x-y)} 
+\frac {W_{\beta^*}(z)}{2i\pi} \\
\ds \frac {(-)^n(2i\pi)^2}{h_{n-1}} p_{n-1}(z) & \ds\frac {(-)^n 2i\pi}{h_{n-1}} \!\! \int_{\R_+} \!\!\!\!\frac {p_{n-1}(x) {\rm e}^{-V_{_1}(x)}\d x}{x-z} & \ds \frac {(-)^n} {h_{n-1}} \int_{\R_+}\int_{\R_-} \frac {p_{n-1}(x) {\rm e}^{-NV_{_1}(x) -NV_{_2}(y)}\d x\d y}{(y-z)(x-y)} 
\end{bmatrix}\nonumber
\eea
where $\wh p_n$ are certain polynomials of exact degree $n$ described in \cite{Bertola:CBOPs}, $W_{\beta^*}(z)=\int_{\R_-}\frac{{\rm e}^{-NV_2(y)}}{y-z}\d y $ and all integrals 
are oriented integrals.
We will study the asymptotic behavior of the solution of Problem \ref{RHPp} in various regions of the complex plane for 
\be
\N \ni N\to\infty\ ,\ \ n:= N+r, \ r\in \Z\ ,
\ee
where the integer $r$ is bounded.
\section{The $\gg$ functions}
\label{seclargen}

An asymptotic treatment  for the RHPs for $\Gamma$ along the lines of the nonlinear steepest descent method \cite{DKMVZ}, requires that we {\em normalize} the problem with the use of an auxiliary matrix constructed from {\bf equilibrium measures} that  minimize a functional described below (see \cite{BertolaBalogh},\cite{Bertola:CauchyMM}).

Consider the space of pairs of probability measures $\mu_1, \mu_2$ supported on $\R_+$ ,$\R_-$ respectively.  On this space we define the functional
\bea
S[\d \mu_1,\d \mu_2] := \int_{\R_+} V_1(x)\d \mu_1(x)
+  \int_{\R_+} \int_{\R_+} \d \mu_1(x)\d \mu_1(x') \ln \frac 1{|x-x'|}  +\cr  + \int_{\R_-} V_2(y) \d \mu_2(y)  
+  \int_{\R_-} \int_{\R_-} \d \mu_2(y)\d \mu_2(y') \ln \frac 1{|y-y'|}  +\cr   \int_{\R_+} \int_{\R_-} \d \mu_1(x)\d \mu_2(y) \ln |x-y| \label{functional}
\eea

The minimization of such a functional was studied in a more general setting in \cite{BertolaBalogh}, it is related to a similar problem for a vector of measures of Nikishin type \cite{VanAssche:RHPMOPs}.
\bt {\rm {(see Theorem 3.2 in \cite{BertolaBalogh})}}

\begin{wrapfigure}{r}{0.7\textwidth}
\resizebox{0.64\textwidth}{!}{
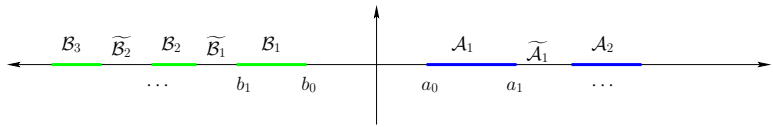}
\caption{The supports of the equilibrium measures $\rho_1, \rho_2$}
\label{rhosupp}
\end{wrapfigure}

\label{Propprop}
Under the Assumptions \ref{ass1} there exists a unique pair of densities $\rho_1, \rho_2$ that minimizes the functional (\ref{functional}). Moreover (see Fig. \ref{rhosupp})
the supports consist of a {\bf finite union} of compact intervals and the densities $\rho_j$ are smooth on the respective supports;
\bea
\supp(\rho_1) = \bigsqcup_{\ell=1}^{L_1}\mathcal A_\ell \subset \R_+\ ,\qquad 
\supp(\rho_2)  = \bigsqcup_{\ell=1}^{L_2}  \mathcal B_\ell \subset \R_-\ .
\eea
\et
\parbox[t]{0.4\textwidth}{  }
\br
\label{hardedge}
It has been recently proven by one of the authors and A. Kuijlaars that the growth condition near the origin in  (\ref{growthcondV}) can be simply disposed of. The properties in Theorem \ref{Propprop} are still valid except that the support of equilibrium measures may contain 
$x=0$, in which case $\rho_j(x) = \mathcal O(x^{-\frac 2 3})$. This behavior is crucial in deriving a new type of universality near $x=0$ and will be part of a forthcoming publication. 
\er

\bt{\rm {(Theorem 5.1 in \cite{Bertola:CauchyMM}, see also \cite{BertolaBalogh})}}
\label{speccurve}

The {\bf shifted resolvents}
\begin{align}
Y^{(1)}:= -W_1 + \frac {2V_1'+V_2'}{3}\ ,&\qquad
Y^{(2)}:= W_2 - \frac {V_1'+2V_2'}{3}, \text{ where } W_j(z):= \int \frac {\rho_j(x)}{x-z} \d x,\nonumber  \\
Y^{(1)} + &Y^{(2)} + Y^{(0)} = 0\ .\label{shiftresolvents}
\end{align}
 are the three branches of the same {\bf cubic} equation in the form 
\be
E(y,z) := y^3 - R(z) y - D(z) = 0\ , \label{algcurve}
\ee
where $R(z), D(z)$ are certain functions analytic in the common domain of $V_1'$ and $V_2'$.  
\et
As a corollary of Theorem \ref{speccurve} we deduce the
\bc
\label{squareroot}
For generic real analytic potentials the densities of the two equilibrium measures $\rho_1,\rho_2$ vanish like a square root at the endpoints of each interval $\mathcal{A}_{\ell}, \mathcal{B}_{\ell}$ 
in the support of the spectrum.  
\ec
{\bf Proof.} It follows immediately from Cardano formul\ae, since the densities are related by Theorem \ref{speccurve} to the jump-discontinuity of the branches of the pseudo--algebraic curve (\ref{algcurve}) which have in general square-root type singularities corresponding to simple zeroes of the discriminant $\Delta=4R^3(z)-27D^2(z)$.
\QED

\begin{figure}
\resizebox{1\textwidth}{!}{
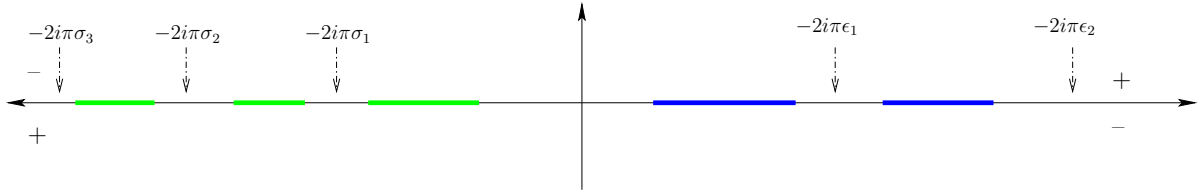}
\caption{The jumps of the $\gg^{(1)}$ and $\gg^{(2)}$ functions in the gaps.}
\label{figGjumps}
\end{figure}
\bd[$\gg$--functions]
\label{defgg}
The $\gg$--functions are defined as 
\be
\gg^{(1)}(z):= \int_{\R_+}\!\!\! \rho_1(x) \ln (z-x) \d x\ ;\  \gg^{(2)}(z):= \int_{\R_-} \!\!\!\rho_2(x) \ln (z-x)\d x\ ;\ 
\gg^{(0)}(z) + \gg^{(1)}(z) + \gg^{(2)}(z) \equiv 0
\ee
where $\gg^{(1)}$ is defined as an analytic function in the domain $\DD_1:=\C\setminus [a_0,\infty)$ and $\gg^{(2)}$ is analytic in $\DD_2:= \C\setminus (-\infty, b_0]$ while $\gg^{(0)}$ is analytic in $\DD_0:= \DD_1\cap \DD_2 = \C \setminus \le((-\infty, b_0] \cup [a_0,\infty) \ri)$
\ed
\br
We remind the reader that the integral defining $\gg^{(2)}$ is an oriented integral.
The orientation of the negative half-axis that we use may seem unusual. However, it is consistent with the one used in studies of multiple orthogonal polynomials on a collection of radial rays in the complex plane.
\er

\bd
The right and left cumulative filling fractions are defined as: 
\bea
\epsilon_\ell :=\int_{a_0} ^{a_{2\ell -1}}\rho_1(w)\d w, \qquad  \ell = 1,\dots\label{cumulativeepsilon}\\
\sigma_\ell := \int_{b_0}^{b_{2\ell -1}} \rho_2(w)\d w, \qquad \ell = 1,\dots\label{cumulativesigma}
\eea
\ed
Note that $\epsilon_{L_1}=1$ while $\sigma_{L_2}=-1$.  

The variational inequalities that characterize the equilibrium measures $\rho_j$ \cite{SaffTotik}, together with the Definition \ref{defgg} translate into the following theorem. 
\bt[Analyticity properties for the $\gg$ functions] 
\label{anal_thm}
The following properties hold
\begin{enumerate}
\item $\gg^{(1)}$ is analytic in $\DD_1$ and has the asymptotic behavior $\gg^{(1)} = \ln z + \mathcal O(z^{-1}), \quad z\rightarrow \infty$;
\item $\gg^{(2)}$ is analytic in $\DD_2$ and has the asymptotic behavior $\gg^{(2)} =  -\ln z + \mathcal O(z^{-1}), \quad z\rightarrow \infty$;
\item $\gg^{(0)}$ is analytic in $\DD_0$ and has the asymptotic behavior $\gg^{(0)} =  \mathcal O(z^{-1}), \quad z\rightarrow \infty$;

\item on the {\bf right cuts} $\mathcal A_\ell$
the function $\Im \le(\gg_+^{(1)}(x) - \gg_-^{(1)}(x)\ri)$ is decreasing and there exists a constant $\gamma_+$ such that 
\be
\gg_\pm^{(0)}(x) - \gg_\mp^{(1)}(x) +V_1(x) +\gamma_+ \equiv 0
\label{1jmp}
\ee

\item on the {\bf right gaps} $\wt{\mathcal A}_\ell:$ 
\bea
\gg^{(1)}_+(x) - \gg^{(1)}_-(x) = -2i\pi\epsilon_\ell =\hbox{ constant }\in i\R, \\
\Re \le(\gg_+^{(0)}(x) - \gg_-^{(1)}(x) +V_1(x)   +\gamma_+\ri) \geq  0,\label{eff1}
\eea

\item on the {\bf left cuts} ${\mathcal B}_{\ell} $
the function $\Im \le(\gg_+^{(2)}(x) - \gg_-^{(2)}(x)\ri) $ is decreasing and there 
exists a constant $\gamma_-$ such that 
\be 
\gg_\pm^{(2)}(x) - \gg_\mp ^{(0)}(x) +V_2(x) +\gamma_-\equiv 0;
\label{2jmp}
\ee

\item on the {\bf left gaps} $\wt{\mathcal B}_\ell $: 
\bea
\gg^{(2)}_+(x) - \gg^{(2)}_-(x) = -2i\pi \sigma_\ell =\hbox{ constant }\in i\R, \\
\Re \le(\gg_+^{(2)}(x) - \gg_-^{(0)}(x) +V_2(x) +\gamma_- \ri) \geq  0, 
\label{eff2}
\eea

\item  on the gaps $\wt{\mathcal B}_\ell\bigcup \wt {\mathcal A}_\ell $:
\be
\gg^{(0)}_+(x) - \gg^{(0)}_-(x) = \le\{\begin{array}{cc}
2i\pi \epsilon_\ell & x \in \wt{\mathcal A}_\ell \\
2i\pi \sigma_\ell & x \in \wt{\mathcal B}_\ell 
\end{array}\ri.
\ee
\end{enumerate}

\et
The proof is based on the facts that $\rho_j$ are {\em positive} densities, the variational (in)equalities that arise from the minimization of the functional (\ref{functional}) \cite{SaffTotik, BertolaBalogh} and elementary complex function theory.  We leave the details to 
the reader.  
\bd \label{regular}
If the statement that functions 
in items 4. and 6. above are decreasing 
 is replaced by the requirement that their derivatives are strictly negative the inequalities in (\ref{eff1}, \ref{eff2}) are strict then the potentials are said to be {\bf regular}.
\ed
Because of the expressions (\ref{eff1}, \ref{eff2}) we make the following 
\bd
\label{effdef}
Define the {\bf effective complex potentials} by
\bea
\varphi_{_1}(z) := V_1(z)  -\gg^{(1)}(z)+\gg^{(0)}(z) + \gamma_+, \\
\varphi_{_2}(z) := V_2 (z)  -\gg^{(0)}(z)+\gg^{(2)}(z) + \gamma_-. 
\eea
\ed
The form of the $\gg$--functions (Definition \ref{defgg}) implies some important inequalities and equalities for the effective potentials that we presently explore.

\bt
\label{effsign}
The effective complex potentials $\varphi_{_1}, \varphi_{_2}$  satisfy the following properties
\begin{enumerate}
\item For any $\ell$ there exists an open neighborhood of $(a_{2\ell -2},a_{2\ell -1})$  for which 
the real part of $\varphi_{_1}$ is {\em negative} away from the cut $\mathcal A_{\ell} $;
\item For any $\ell$ there exists an open neighborhood of $(b_{2\ell -2},b_{2\ell -1})$  for which the real part of $\varphi_{_2}$  is {\em negative} away from the cut $\mathcal B_\ell$.
\end{enumerate}
\et 
{\bf Proof}.
It suffices to consider any of the right cuts, say, $\mathcal A_\ell$.  Then because of the identity $\sum \gg^{(j)}\equiv 0$  and from the properties specified in Theorem \ref{anal_thm}  (since $\gg^{(2)}_+ - \gg^{(2)}_- =0$) we have
\bea
\gg^{(1)}_+ -\gg^{(1)}_- = -\gg^{(0)}_+ +\gg^{(0)}_- 
\ \ \Rightarrow\ \ 
\gg^{(1)}_+ +\gg^{(0)}_+ = \gg^{(1)}_- +\gg^{(0)}_- . 
\eea
More importantly, we see from (\ref{1jmp}) above that 
\bea
\gg^{(1)}_+ - \gg^{(1)}_- &\& = \gg^{(1)}_+ - \gg ^{(0)}_+  -V_1  - \gamma_+ = -\varphi_{{1}+}=\label{leftBV}\\
&\& = -\le (\gg^{(1)}_- - \gg^{(0)}_- -V_1 -\gamma_+ \ri) = \varphi_{{1}-}\label{rightBV}
\eea
Equation \eqref{leftBV} implies that the jump $\Delta \gg^{(1)}$ is the left boundary value of the  analytic function $-\varphi_1$, while equation \eqref{rightBV} represents the same quantity as the right boundary value of the  analytic function $\varphi_1$.  The condition  of strict decrease appearing in Theorem \ref{anal_thm} then implies in view of the Cauchy--Riemann equations  that 
$ \Re \varphi_{{_1}}$ strictly decreases as one moves perpendicularly away from the cut. \QED

Since the three $\gg$--functions are antiderivatives of the resolvents $W_j$ \eqref{shiftresolvents} we deduce that there must be two constants $\gamma_{1,2}$ such that  
\be
\gg^{(1)}(z)=  \gamma_1 +\frac {2V_1+V_2}{3}-\int_{a_0}^z Y^{(1)}(\zeta) \d \zeta \ ; \ \ 
\gg^{(2)}(z) = \gamma_2 -\frac{2V_2+V_1}{3}-\int_{b_0}^z Y^{(2)}(\zeta)\d \zeta  \ ,\label{Yrep}
\ee
where the respective integrals are performed within the simply connected domains specified above and the constants $\gamma_{1,2}$ are chosen so that asymptotically 
\bea
\gg^{(1)}(z) = \ln z + \mathcal O(z^{-1}),   \qquad \ \gg^{(2)}(z) = -\ln z + \mathcal O(z^{-1}). 
\eea
It thus appears that the three $\gg$--functions are intimately related to the three branches of the integral $\int y \d z$ over the three--sheeted Riemann surface (\ref{algcurve}).  
\section{Deift--Zhou steepest descent analysis}
\label{DZdescent}

We will follow a well-established approach to asymptotic analysis of Riemann-Hilbert problems pioneered by Deift and Zhou
in \cite{DeiftZhou} and referred to as the {\em nonlinear steepest descent  method}.

\subsection{ Modifications of the Riemann-Hilbert Problem} 
\begin{figure}[h]
\resizebox{0.9\textwidth}{!}{
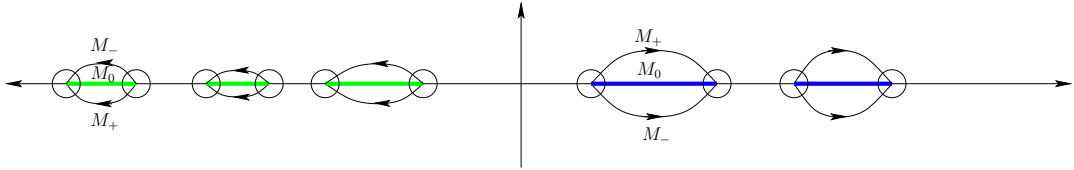
}
\caption{The final steps in the modification of the RHP.}
\label{CutsLens2fig}
\end{figure}

We proceed to introduce the contours indicated in Fig. \ref{CutsLens2fig}:  the disks $\mathbb D_a$  around each endpoint $a\in \{ a_j, b_j\}_{j=1,\dots}$ of the cuts  have sufficiently small radii so as not to include any other end points. The exact position and shape of the  upper and lower arcs joining two end  points are largely irrelevant; they should lie in the region where the real parts of the effective potentials (Definition \ref{effdef}) are negative.
We will name  the part outside all disks and all lenses the {\bf outer region}. 

Let us define a new piecewise analytic matrix function (only nonzero entries are indicated)

\begin{figure}[h]
\resizebox{0.8\textwidth}{!}{
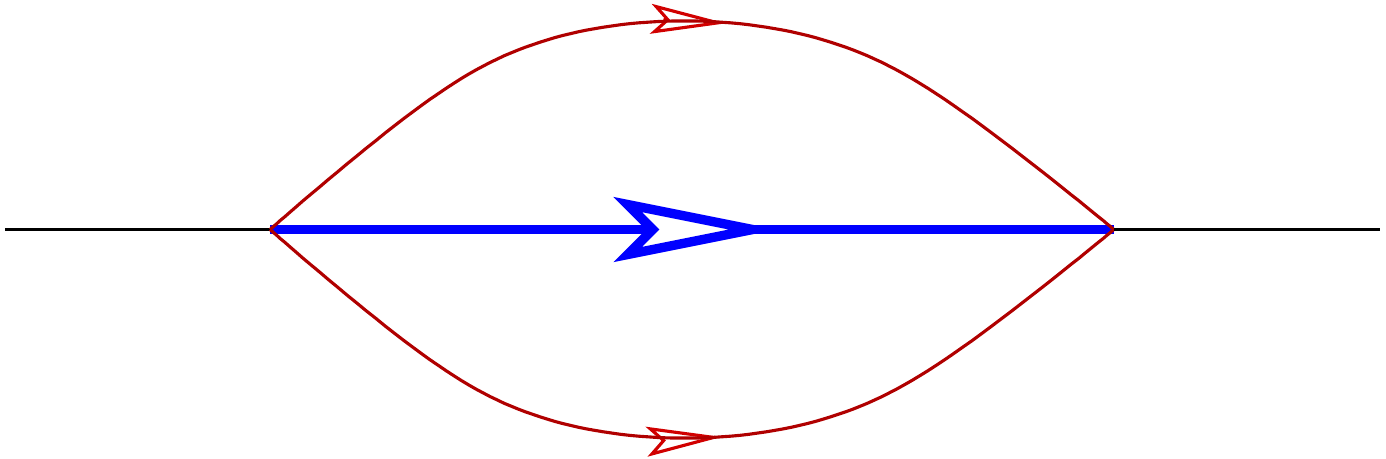}
\caption{The modified jumps for $ \Gamma_{_0}$ on a cut of the  {\em right} side of the spectrum.}
\label{FigGamma0}
\end{figure}
\begin{figure}
\resizebox{0.8\textwidth}{!}{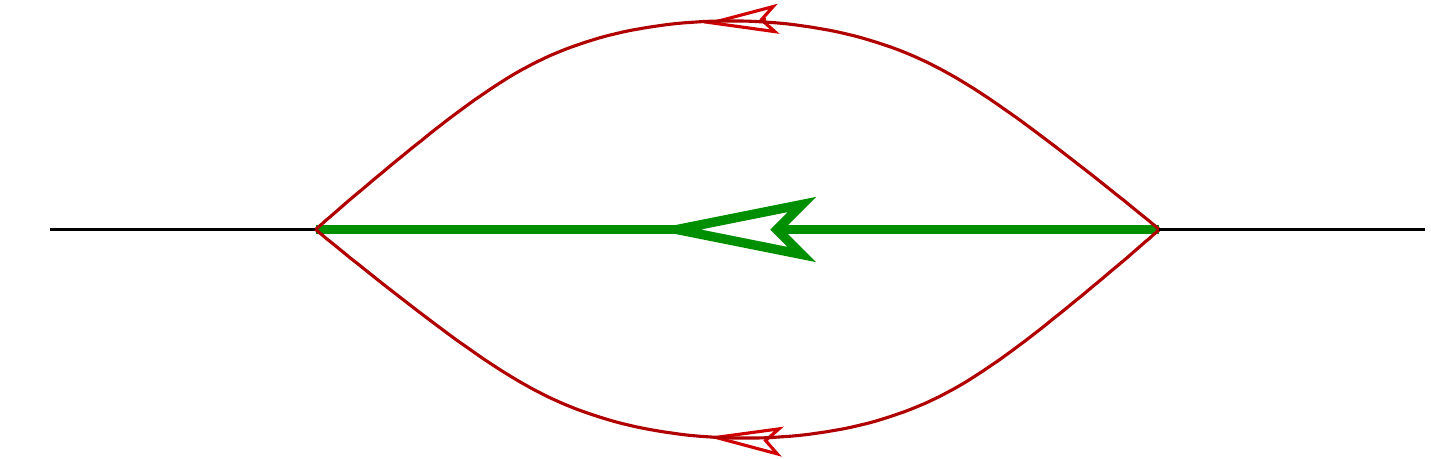}
\caption{The modified jumps for $ \Gamma_{_0}$ on a cut of the {\em left} side of the spectrum.}
\label{FigGamma0left}
\end{figure}
\bea
 \Gamma_{_0} := \le\{
\begin{array}{cc}
  \Gamma & \hbox{ in the outer region}\\[10pt]
  \Gamma \le[
 \begin{array}{ccc}
 1&&\\
 -{\rm e}^{N V_1(z)} & 1 & \\
 &&1
 \end{array}
 \ri] & \hbox{ in the upper half of each lens on the right cuts}\\[3pt]
  \Gamma \le[
 \begin{array}{ccc}
 1&&\\
 {\rm e}^{N V_1(z)} & 1 & \\
 &&1
 \end{array}
 \ri]  & \hbox{ in the lower half of each lens on the right cuts}\\[10pt]
   \Gamma \le[
 \begin{array}{ccc}
 1&&\\
 &1&\\
  &{\rm e}^{N V_2(z)}&1
 \end{array}
 \ri] & \hbox{ in the upper half of each lens on the left cuts}\\[10pt]
  \Gamma \le[
 \begin{array}{ccc}
 1&&\\
& 1 & \\
 & - {\rm e}^{N V_2(z)} &1
 \end{array}
 \ri]  & \hbox{ in the lower half of each lens on the left cuts}\\
\end{array}
\ri.
\eea 
As a consequence the jumps of the Riemann--Hilbert problem are modified as indicated by the jump--matrices in Figs. \ref{FigGamma0}, \ref{FigGamma0left}. 
We now define 
\bea
 \ZZ (z)&\& := C_\gamma^{-1}  \Gamma_{_0}(z)G(z) C_{\gamma}\ ,\qquad G(z):= {\rm diag} ({\rm e}^{-N\gg^{(1)}(z)} ,{\rm e}^{ -N \gg^{(0)}(z)},{\rm e}^{ -N\gg^{(2)}(z)})
\label{maintransform}
\\
&\& C_\gamma  := \le[
\begin{array}{ccc}
\ds{\rm e}^ { N \frac {\gamma_- + 2\gamma_+}{3}} & &\cr
& {\rm e}^{ N \frac {\gamma_- - \gamma_+}{3}} & \cr
&& {\rm e}^{- N \frac {\gamma_+ + 2\gamma_-}{3}}
\end{array}
\ri]
\eea
The matrix $\ZZ$  will be the  main object of our interest from this
point on. Recall that $\gg ^{(j)}=(-1)^{j} \ln z + \mathcal{O}(z^{-1}), \text{ for } j=1,2, $ and that $n=N+r$.  Then 
$\ZZ$ satisfies a new Riemann--Hilbert problem:
\be
\ZZ(z)  = \le(\1 + \mathcal O(z^{-1})\ri) z^{ {\rm diag}(r,0,-r)}\ \ \ \ z\to\infty,\qquad
 \ZZ_+(z) = \ZZ_-(z) M(z) 
\ee
where $M(z)$ takes different forms depending on the arc considered. 
Note that $\ZZ$ is bounded at $z=0$ because of Assumption \ref{ass1} and Remark \ref{strongg}.
Recalling the definition of the effective potentials (Def. \ref{effdef}) we have (using the notation $\Delta f = f_+-f_-,\ \  \mathcal S f = f_++f_-$) 
\bea
\begin{split}
M = M_{\pm}^{(right)} := \le[ 
\begin{array}{ccc}
1&&\\
{\rm e}^{N\varphi_{_1}}&1\\&&1
\end{array}
\ri],  \ \hbox { on the upper/lower rims of a right lens}, \\
M = M_{\pm}^{(left)} := \le[ 
\begin{array}{ccc}
1&&\\
&1&\\
& {\rm e}^{N\varphi_{_2}}&1
\end{array}
\ri], \ 
\hbox{ on the upper/lower rims of a left lens}, 
\end{split}
\label{Mlens}
\eea

\bea
M = M_0^{(right)} := \le[
\begin{array}{ccc}
& {\rm e}^{-N\le( V_1  -\gg^{(1)}_- + \gg_+^{(0)} + \gamma_+ \ri)}&\\
 -{\rm e}^{N\le( V_1 -\gg^{(1)}_- + \gg_+^{(0)}+ \gamma_+  \ri)} &&\\
 &&1
\end{array}
\ri] = \le[
\begin{array}{ccc}
&1&\\
 -1 &&\\
 &&1
\end{array}
\ri], 
\eea
on the right cuts, 
\bea
\begin{split}
M = M_0^{(left)} := \le[
\begin{array}{ccc}
1&&\\
&& {\rm e}^{-N\le(V_2  -\gg^{(0)}_- + \gg_+^{(2)} + \gamma_- \ri)}\\
&-  {\rm e}^{N\le( V_2 -\gg^{(0)}_- + \gg_+^{(2)}+ \gamma_-  \ri)} &
\end{array}
\ri] = \le[
\begin{array}{ccc}
1&&\\
&&1\\
 &- 1&
\end{array}
\ri], 
\end{split} \label{Mcut}
\eea
on the left cuts, 
\bea
\begin{split}
M= M_{gap}^{(right)} =  \le[
\begin{array}{ccc}
{\rm e}^{2i\pi N \epsilon_\ell} &{\rm e}^{N\le(\gg_+^{(0)}- \gg_-^{(1)} -  V_1-\gamma_+\ri)}&\\
&{\rm e}^{-2i\pi N \epsilon_\ell}&\\
&&1
\end{array}
\ri] =  \le[
\begin{array}{ccc}
{\rm e}^{\frac  N2 \Delta \varphi_{_1}} &{\rm e}^{-\frac N2\mathcal S\,\varphi_{_1 } }&\\
&{\rm e}^{- \frac N2 \Delta\varphi_{_1 } }&\\
&&1
\end{array}
\ri]
\end{split} 
\label{Mrgap}
\eea
on the right gaps, 
\bea
\begin{split}
M= M_{gap}^{(left)} =  \le[
\begin{array}{ccc}
1&&\\
&{\rm e}^{-2i\pi N \sigma_\ell} &{\rm e}^{N\le(\gg_+^{(2)}- \gg_-^{(0)} - V_2-\gamma_-\ri)} \\
&&{\rm e}^{2i\pi N \sigma_\ell}
\end{array}
\ri] = 
 \le[
\begin{array}{ccc}
1&&\\
&{\rm e}^{\frac N2\Delta\varphi_{_{2}} } &{\rm e}^{-\frac N2 \mathcal S \varphi_{_{2}}} \\
&&{\rm e}^{-\frac N2\Delta\varphi_{_{2}}}
\end{array}
\ri]
\end{split}
\label{Mlgap}
\eea
on the left gaps.

\begin{figure}
\resizebox{1\textwidth}{!}{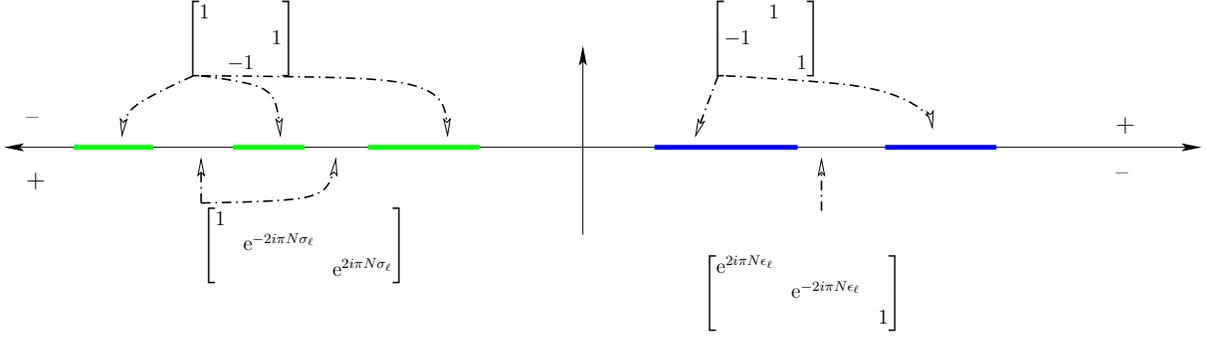}
\caption{The RHP for the outer parametrix}
\label{outerparametrix}
\end{figure}

Observe that the off--diagonal entries of \eqref{Mlens}, \eqref{Mlgap}, and \eqref{Mrgap}  are {\bf
  exponentially small} as $N\to\infty$ because of the signs of the
real parts; in particular all their $L^p$  norms (except for $p=\infty$) are exponentially
small.
Yet, near the endpoints $a_j, b_j$ of the support of the equilibrium measures, the off--diagonal entries tend to $1$. However, the new RHP that effectively amounts to setting them to zero is  a key ingredient in the final approximation we need. It is, in fact, the main new ingredient of the paper from the point of view of the nonlinear steepest descent method. More explicitly, we formulate
\begin{problem}[Outer parametrix]
\label{outerRHP}
Find a $3\times 3$ matrix $\Psi(z)$, analytic in $\mathcal D_0:= \C\setminus \le((-\infty, b_0]\cup [ a_0,\infty)\ri)$ and with the following properties
\begin{enumerate}
\item [(i)] the jumps indicated in  Figure \ref{outerparametrix};
\item[(ii)]  the growth conditions at $z=\infty$ and near an endpoint $z=a$ are, respectively
\be
\Psi(z) = \le(\1 + \mathcal O \le(\frac 1  z \ri)\ri) \begin{pmatrix} z^r  &&\\ & 1& \\ && z^{-r} \end{pmatrix}\ ,\  \Psi(z) = \mathcal O\le((z-a)^{-\frac 1 4}\ri)\ ,\ a \in \{a_i, b_i\}_{i=1,\dots}\label{growthcond}
\ee
\end{enumerate}
\end{problem}

The final approximation involves the solution of the exact RHP for $\ZZ$ within disks around the endpoints that we have indicated above (see Figure \ref{CutsLens2fig}).  This is called the {\bf local RHP} and is formulated below
\begin{problem}[Local parametrix]
\label{localRHP}
Let $\mathbb D = \mathbb D_a$  be a disk (previously introduced) around any of the endpoints $a$ of the supports of $\rho_1,\rho_2$.  Find a piecewise analytic $3\times 3$ matrix $\mathcal P(z) = \mathcal P_a(z)$ such that 
\begin{enumerate}
\item [(i)] $\mathcal P(z)$ is bounded for  $z\in \mathbb D$ uniformly with respect to $N$;
\item [(ii)] within $\mathbb D$, $\mathcal P(z)$ satisfies  the jump-conditions for $\ZZ$ with jump-matrices (\ref{Mlgap}, \ref{Mrgap}, \ref{Mcut}, \ref{Mlens});
\item [(iii)] on the boundary $\pa \mathbb D$,  $\mathcal P(z) \Psi^{-1}(z) = \1 +  o(1), \quad N\rightarrow \infty$, uniformly in $z\in \pa \mathbb D$.
\end{enumerate} 
\end{problem}
Suppose now that we have managed to solve Problems \ref{outerRHP}, \ref{localRHP} and let us define the matrix
\be
\wh \ZZ(z) := \le\{
\begin{array} {cc}
\Psi(z) & z\not\in \bigcup_{a} \mathbb D_a\\
\mathcal P_a(z) &  z\in \mathbb D_a
\end{array}
\ri.\ \ \label{whGamma}
\ee
\begin{figure}
\resizebox{1\textwidth}{!}{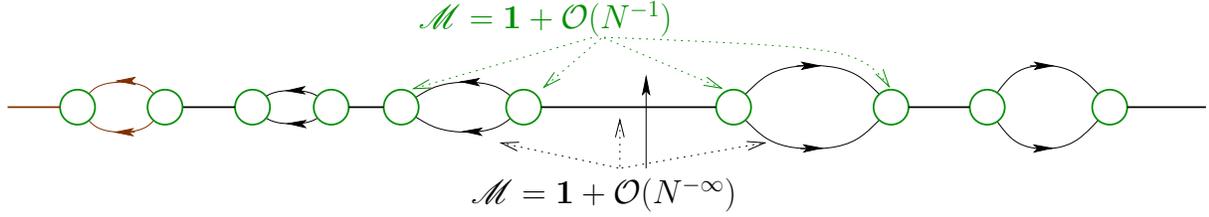}
\caption{The contours of the residual RHP for the error term and the orders of the jumps matrices.}
\label{residual}
\end{figure}

Then the ``error'' matrix $\mathcal E(z):= \ZZ(z) \wh \ZZ^{-1}(z)$ solves the RHP
\be
\mathcal E_+ = \mathcal E_- (\1 + G)\ ,\ \ \ \mathcal E(z) = \1 + \mathcal O(z^{-1})\ ,
\ee
where the jumps are supported on the boundaries $\pa \mathbb D_a$, on the rims of the lenses and gaps {\em outside} the local disks.
A direct standard inspection {( see, e.g. \cite{DKMVZ})} shows that $G$ tends to zero 
as $N\rightarrow \infty$ in all $L^p$ norms $p\in [1,\infty]$ and hence $\mathcal E(z)$ is close to the identity matrix $\1$ uniformly on $\mathbb P^1$. If the equilibrium problem is {\em regular} (in the sense of Remark \ref{regular}) then the bounds on the jumps of the error term $\mathcal E$ (and hence on the error itself) are as depicted in Fig. \ref{residual}. 

 \subsection{Outer parametrix}
The construction of the solution to the RHP \ref{outerRHP} uses Theta-Functions associated to an algebraic curve studied below.  Solutions to problems of this kind can be derived in several ways \cite{Korotkin05, BertolaAdvances}. While we will eventually write down ``explicit" formul\ae\ involving Theta functions, these formul\ae\ are not numerically effective
unless the underlying Riemann surface has genus zero.   
However, we do need to prove {\em existence} of a solution. Thus, our strategy will be to produce theta--functional expressions for the solution to Problem \ref{outerRHP} and subsequently use results on bordered Riemann surfaces from \cite{Faybook} to ensure solvability of the RHP in terms of the proposed expressions.
As an example, we will also provide the corresponding (explicit) expressions for the case of genus zero.
 \subsubsection{The abstract  Riemann surface}
\label{sectAbstractCurve}
The title of this section refers not to the Riemann surface defined by the pseudo-algebraic equation $E(y,z) =0$ (\ref{algcurve}), but to an abstract Riemann surface $\L$ described below.

Let $L_1, L_2$ be the number of intervals of the supports of the measures $\rho_1, \rho_2$; note that endpoints of cuts correspond to zeroes of odd multiplicity of the discriminant equation $
\Delta (z):= 4R^3(z) -27 D^2(z)$ ,
and the assumption of {regularity} of the potentials translates into the requirement that all zeroes of odd multiplicity  of $\Delta(z)$ are real and simple. 

We consider now the abstract Riemann surface obtained by gluing together  three Riemann spheres
parametrized by the variable $z$; the Riemann sphere labelled $1$  is slit along the support of $\rho_1$ and glued there with the middle Riemann sphere labelled $0$. The latter is subsequently slit also along the support of $\rho_2$ and glued across it with the Riemann sphere labelled $2$, cut along the support of $\rho_2$.

The resulting Riemann surface $\mathcal{L}$ is a compact surface of genus 
\be
g=genus(\L) = L_1+L_2-2
\ee
as follows from the Riemann--Hurwitz formula. The endpoints of the intervals $\mathcal A_\ell,\ \ell=1, \dots L_1$ and  $\mathcal B_\ell,\ \ell=1,\dots, L_2$ are branch points of order $2$,  i.e. the local coordinate is $\sqrt{z-x_0}$, with $x_0$ one of the endpoints. The local coordinate around the three points at infinity $\infty_1, \infty_0,\infty_2$ is $1/z$.

\begin{figure}
\resizebox{0.7\textwidth}{!}{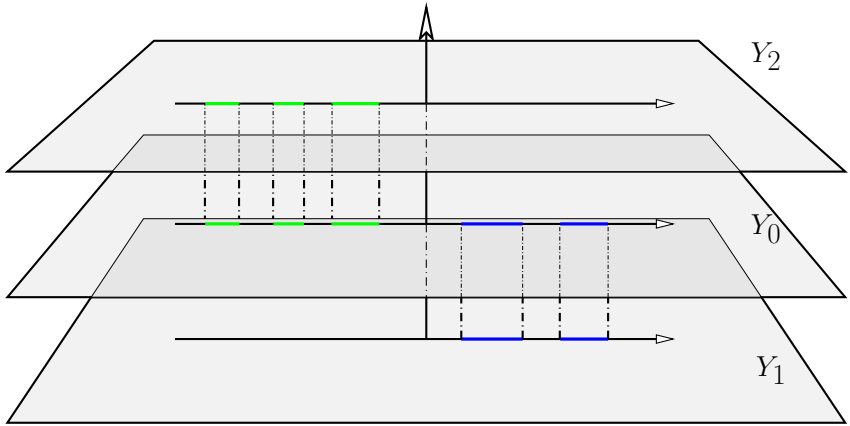}
\caption{The Hurwitz diagram of the abstract spectral curve $\L$: the vertical dotted lines represent the identification of points.}
\label{Hurwitzfig}
\end{figure}

This point of view allows one to think of $Y^{(0,1,2)}$ appearing in Theorem 
\ref{speccurve} as the three branches of a locally analytic function $y$ over the Riemann surface $\mathcal{L}$; indeed near a branch point $x_0$, due to the assumption about $\Delta$ and due to Cardano formulae, the function $y$ has a square-root Puiseux expansion in $z-x_0$, and thus is a well defined function on the curve $\L$.

If $V_1', V_2'$ are meromorphic functions then $y$ is a meromorphic function on $\L$,   otherwise $y$ will  in general have other isolated singularities or will be defined only in a strip around the three--copies of the real axis in $\L$.

\begin{figure}[h]
\resizebox{1\textwidth}{!}{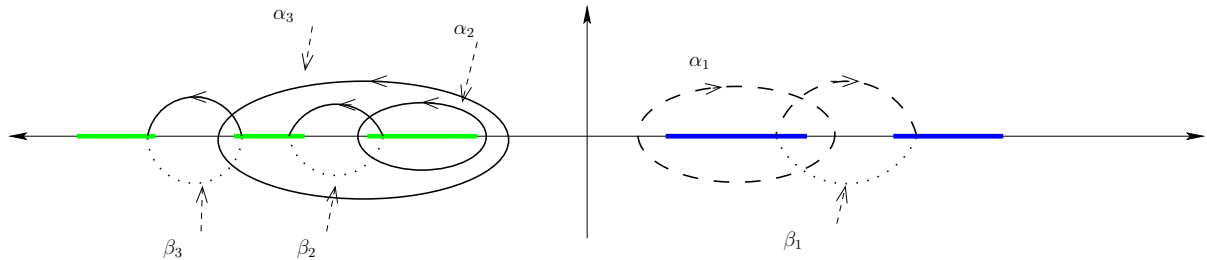}
\caption{Our choice of the canonical homology basis on the abstract Riemann surface $\L$; the solid lines represent arcs on Sheet $2$, the dotted ones are arcs on Sheet $0$, and the dashed ones are arcs on Sheet $1$. In the example there are $5$ total cuts and hence the genus is $3$. We can declare that the curves lying on the same sheet are the $\alpha$--cycles whereas the curves lying on two different sheets are the corresponding $\beta$--cycles.}
\label{Homologyfig}
\end{figure}

We will use the basis of the canonical homology of $\L$ indicated in Fig. \ref{Homologyfig}.

\bp
\label{Mcurve}
The curve $\mathcal{L}$ possesses a natural {\em antiholomorphic
  involution} defined as the complex  conjugation of each of the three sheets. The finite gaps provide $g=genus (\mathcal L)$ point-wise invariant nontrivial cycles. Thus $\mathcal L$ can be realized as a double of the bordered Riemann surface obtained by identifying the upper half of sheet $2$ with the lower half of sheet $0$ along the left cuts, and the lower half of sheet $0$ with the upper half of sheet $1$ along the right cuts (see Chapter 6 in \cite{Faybook}).
\ep
Proposition \ref{Mcurve} shows that the properties of $\mathcal L$ are very similar to the properties of a hyperelliptic curve defined as $w^2 = P(z)$ for a {\em real} polynomial $P$ of degree $2g+2$.  This fact will be used to obtain information on the Theta divisor of the Jacobian of $\mathcal{L} $ when constructing the outer parametrix. 
\subsubsection{Multiplier system $\chi$ and spinors}
\label{DomainD}
Recall the definition of domains $\DD_{0,1,2}$ (Def. \ref{defgg}). By lifting 
points in the complex plane $z$ one can construct 
three sections $p_{j}:\mathcal D_j \to \L$ such that: $p_1$ is analytic in $\DD_1 = \C\setminus \sqcup \mathcal A_\ell$, $p_2$ is analytic in $\DD_2 = \C\setminus\sqcup  \mathcal B_\ell$ and $p_0$ is a analytic in $\DD_0 = \DD_1\cap \DD_2$ with the following boundary values
\bea
&& p_1(x)_\pm = p_0(x)_\mp \ ,\qquad x\in \mathcal A  := \sqcup_{\ell=1}^{L_1}\mathcal A_\ell\cr
&& p_2(x)_ \pm  = p_0(x)_\mp\ ,\qquad x \in \mathcal B:= \sqcup_{\ell=1}^{L_2} \mathcal B_\ell
\eea

\begin{figure}
\resizebox{18cm}{!}{
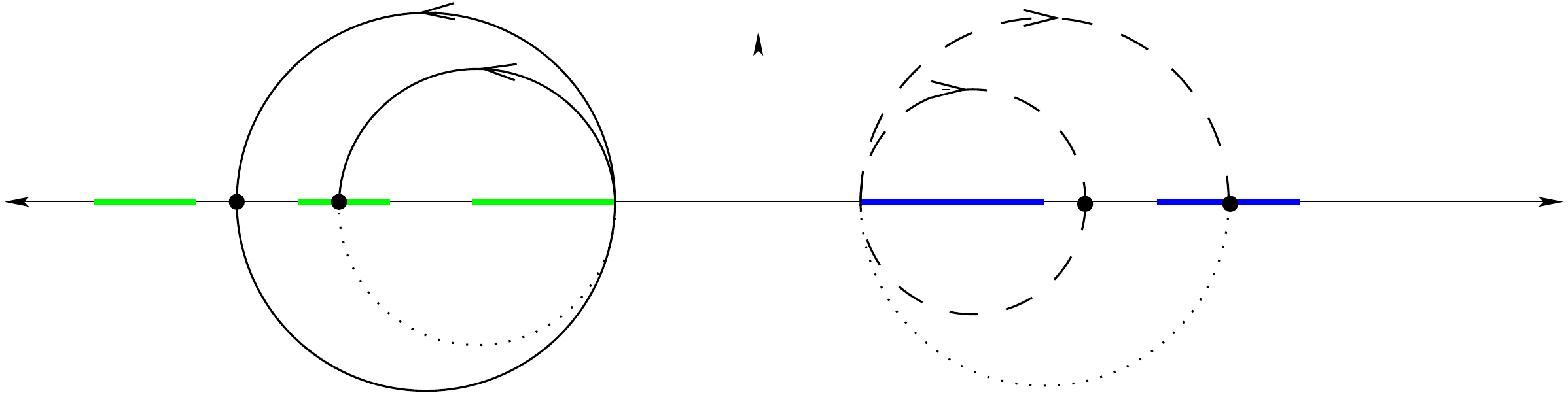}
\caption{Depiction of the lifted contours $\gamma_x,\gamma_{x'}, \gamma_y, \gamma_{y'}$. Their $X$--projection is just a closed path (in the picture we have chosen circles) intersecting the real axis only at $x$ and one of $b_0$ or $a_0$. The points $x', y'$ belong to the gaps and by definition the curves $\gamma_{x'}, \gamma_{y'} $ lie on one sheet and are actually homologic to the $\alpha$--cycles  chosen before (see Fig. \ref{Homologyfig} and {\color{blue} its} caption). }
\label{GammaCurve}
\end{figure}

\bd
Let $x\in\R_+$ ($x\in \R_-$); the contour $\gamma_x: S^1\to \L$ is defined as the unique lift to $\L$ of the closed path on $\C$ that contains the leftmost (rightmost) point $a_0\in \mathcal A$ ($b_0\in \mathcal B$) and intersecting the real axis only at $a_0$ and $x$ (see Fig \ref{GammaCurve}). The lift is accomplished by using the map $p_1:\mathcal D_1\to\L$ for the part in the upper half-plane (or $p_2:\mathcal{D}_2 \to \mathcal{L}$ if $x\in \R_-$). The lower part of the path is lifted using $p_0$ if $x$ belongs to a cut or $p_1$ ($p_2$) when $x$ belongs to a gap.
\ed
 
By this definition  $\gamma_x$ defines a closed cycle in the homology of $\L$.  
Moreover, $\gamma_x$ span the whole homology as $x$ ranges through  $\R$.

We define the following vectors (characteristics) in the Jacobian of the curve $\L$
\be
\mathbf A=2i\pi N \le[\begin{array}{c}
\epsilon_1\\
\vdots\\
\epsilon_{L_1-1}\\
\sigma_{1}\\
\vdots\\
\sigma_{L_2-1}
\end{array}\ri]\ ,\qquad
\mathbf B = 0\in \C^g\ .
\label{characters}
\ee
where the cumulative filling fraction have been introduced in (\ref{cumulativeepsilon}, \ref{cumulativesigma}). Note the linear dependence of the vector $\mathbf A$ on $N$ .

Given any two vectors $\mathbf A=(A_j), \mathbf B=(B_j) \in \C^g$ 
one can define a {\bf character}, namely, a homomorphism $\chi: \pi_1(\mathcal L) \to \C^*$  (with $\pi_1$ the fundamental group of the Riemann surface)
 by extending the values it takes on the basis of the homology,  $
\chi( \alpha_j) = {\rm e}^{A_j}\ ,\ \chi(\beta_j) = {\rm e}^{B_j}$.
For $\mathbf A, \mathbf B$ in \eqref{characters} we have 
\be
\chi(\alpha_j) = \le\{
\begin{array}{cc}
{\rm e}^{N2i\pi \epsilon_j} & j\leq L_1-1\\
{\rm e}^{N2i\pi \sigma_{_{j-L_1+1}}} & j> L_1-1
\end{array}
\ri.\ ,\qquad \chi(\beta_j) = 1.  
\label{chiN}
\ee

A {\bf  meromorphic spinor $\psi$} for the character $\chi$ is a multivalued half-differential that  acquires the factor $\chi(\gamma)$ under analytic continuation along a closed contour $\gamma$. Any half-differential $\phi$ at a point $p$ can be expressed  in a local coordinate $\zeta$ as 
$\phi(p)  = f(\zeta)\sqrt {\d \zeta}$. In particular, since $z$ is a local coordinate away from the branch points, we can write $\psi(p) = f(z) \sqrt {\d z}$, where $p\in \L$ is a preimage of $z$. Recalling the definition of the three sections $p_j(z)$ given above we have
\bea
&& \psi(p_1(x))_{\pm} = \chi(\gamma_x) \psi (p_0(x))_{\mp}\ ,\quad x \in \mathcal A\ ,\quad 
\psi(p_1(x))_+ = \chi(\gamma_x) \psi(p_1(x))_- \ ,\quad x \in \R_+ \setminus \mathcal A
 \cr
&&
 \psi(p_2(x))_{\pm} = \chi(\gamma_x) \psi (p_0(x))_{\mp} \ ,\quad x\in \mathcal B 
 \ ,\quad  \psi (p_2(x))_+ = \chi(\gamma_x) \psi(p_2(x))_- \ ,\quad x\in \R_-\setminus \mathcal B\\
&& \psi(p_0(x))_+ = \chi(\gamma_x)^{-1}  \psi(p_0(x))_- \ ,\quad x \in \R_+ \setminus (\mathcal A\cup \mathcal{B})
 \eea
In addition, we will use the spinor $\sqrt{\d z}$ defined on $\mathcal{L}$ slit along 
 the top of the cuts $\mathcal A$ in sheet $1$ and bottom on sheet $0$, and along the top of $\mathcal B$ in sheet $2$ and bottom in sheet $0$.
By definition $\sqrt{\d z}$ satisfies the boundary conditions
\bea
\sqrt{\d z}(p_1(x))_{\pm} =
\pm 
 \sqrt{\d z}(p_0(x))_{\mp} \ ,\ \ 
\ ,\ \ x\in \mathcal A\label{noncloseA}\\
\sqrt{\d z}(p_1(x))_+ =  \sqrt{\d z}(p_1(x))_-\ \ ,\ \ x\in \R\setminus \mathcal A\\
\sqrt{\d z}(p_2(x))_{\pm} = 
\pm\sqrt{\d z}(p_0(x))_{\mp} \ ,\ \ ,\ \ x\in \mathcal B
\label{noncloseB}\\
\sqrt{\d z}(p_2(x))_+ =  \sqrt{\d z}(p_2(x))_-\ \ ,\ \ x\in \R\setminus \mathcal B
\eea
Note that the jump relations (\ref{noncloseA}, \ref{noncloseB}) imply that $\sqrt{\d z}$ is globally defined on a double cover of $\mathcal L$ branched at the {\em ramification} points (to be distinguished from the branch points) since a $2\pi$--loop around one such point (i.e. a $4\pi$-loop around $x=c$ for $c$ a branch point) yields a transformation $\sqrt{\d z}\to -\sqrt{\d z}$.
If we define a multivalued function $f(p) := \frac {\psi(p)}{\sqrt{\d X(p)}}$ it follows from the above that its composition with the three sections $p_j$ (Sec. \ref{DomainD}) satisfies the following boundary--value conditions
\bea
&& f(p_1(x))_{\pm} =
\pm
 \chi(\gamma_x) f (p_0(x))_{\mp} \ ,\ \ \ 
\qquad x \in \mathcal A\cr
&& f (p_1(x))_+ = \chi(\gamma_x) f(p_1(x))_- \ ,\qquad x \in \R_+ \setminus \mathcal A
 \label{monodromyA}\\
&&
f(p_2(x))_{\pm}  = 
\mp
 \chi(\gamma_x) f (p_0(x))_{\mp}  \ ,\ \ 
 \qquad x\in \mathcal B\cr
 && f (p_2(x))_+ = \chi(\gamma_x)f(p_2(x))_- \ ,\qquad x\in \R_-\setminus \mathcal B.  \label{monodromyB}
 \eea
 
For our choice (\ref{chiN}) of $\chi$ the above relations take the form:
\bea
&& f(p_1(x))_+ = 
  f (p_0(x))_- \ ,\ \ \ 
 f(p_0(x))_+ = -f (p_1(x))_- \ ,\qquad x \in \mathcal A\cr
&& f (p_1(x))_+ = {\rm e}^{2i\pi N \epsilon_j} f(p_1(x))_- \ ,\qquad x \in (a_{2j-1},a_{2j})
 \label{monodromyAbis}\\
&&
f(p_2(x))_+ = 
-
  f (p_0(x))_- \ ,\ \ 
 f(p_0(x))_+ =
    f (p_2(x))_- \ ,\qquad x\in \mathcal B\cr
 && f (p_2(x))_+ = {\rm e}^{2i \pi N \sigma_\ell} f(p_2(x))_- \ ,\qquad x\in (b_{2\ell -1}, b_{2\ell} ).  \label{monodromyBbis}
 \eea
 Moreover, near a branch point $x=a$, since the local coordinate on the curve $\L$ is $\sqrt{z-a}$, the functions $f_j(z):= f(p_j(z))$ behave as
 \be
 f_j(z) =\mathcal O( (z-a)^{-\frac 1 4}).
 \ee
 
 All the above discussion amounts to the following 
 \bp \label{propjumps} 
 Let $\psi$ be a meromorphic spinor for the character $\chi$ and let $f_j(\zeta) := \frac {\psi(p_j(\zeta))}{\sqrt {\d z(p_j(\zeta))}}$.  Then the vector 
 \be
 {\bf F} (z):= [f_1(z),  f_0(z), f_2(z)]\ ,\ \ z\in \C \setminus (-\infty,b_0]\cup [a_0,\infty)
 \ee
 has the properties
 \be
 \begin{array}{cc}
 {\bf F} (x)_+  = {\bf F} (x)_- \begin{pmatrix} & 1 & \\ -1 &&\\ &&1
 \end{pmatrix}\ ,\  x\in \mathcal A ;&  
   {\bf F} (x)_+  = {\bf F} (x)_- \begin{pmatrix}{\rm e}^{2i\pi N \epsilon_j} \!\!&  & \\  &\!\!\!\!\!{\rm e}^{-2i\pi N \epsilon_j}\!\!\!\!\!\!&\\ &&1
 \end{pmatrix}
\ ,\ x\in (a_{2j-1},a_{2j})\\[30pt]
 {\bf F} (x)_+  = {\bf F} (x)_- \begin{pmatrix}1 &  & \\  &&1\\ &-1&
 \end{pmatrix}\ ,\ x\in \mathcal B;
 & 
 {\bf F} (x)_+  = {\bf F} _- \begin{pmatrix}1&  & \\  &\!\!\!{\rm e}^{-2i\pi N \sigma_\ell}\!\!\!\!\!\!&\\ &&{\rm e}^{2i\pi N \sigma_\ell} 
 \end{pmatrix}
\ ,\ x\in (b_{2\ell -1}, b_{2\ell} )
 \end{array}
 \label{nicejumps}
 \ee
 as well as ${\bf F} (x) = \mathcal O( (x-a)^{-\frac 1 4})$ for any endpoint $a$. 
 \ep

This means that the jump conditions for the Problem \ref{outerRHP} are already satisfied.  The complete solution amounts to finding a spinor $\psi$ for each {row} such that it has an appropriate growth at the points above $z=\infty$.  Generically, a spinor for a character $\chi$ is uniquely determined, up to a multiplicative constant, by choosing a divisor of degree $-1$.
This is a consequence of Riemann--Roch--Serre theorem \cite{Gunning, Atiyah}  for line-bundles (our line bundle is the tensor product of a flat line bundle defined by the character $\chi$ and a spinor bundle, namely a line bundle whose square is the canonical bundle)
 We have 
\bp
\label{propbehaviour}
 Consider the two sequences of  spinors $\psi_r, \wh \psi_r,\ r\in \Z$ for the same character $\chi$ (\ref{chiN}) satisfying the following divisor properties (see (\ref{divisornotation}) in the appendix)
\be
(\psi_r) \geq -(r+1)\infty_1 + r\infty_2\ ;\qquad (\wh \psi_r)\geq -(r+1)\infty_1 - \infty_0 +(r+1)\infty_2
\ee
Consider the matrix  
\be
\mathfrak V(z) := \begin{pmatrix}
f_r(p_1) & 
f_r(p_0) &
f_r(p_2) \\ 
\wh f_{r-1}(p_1) &
\wh f_{r-1}(p_0) &
\wh f_{r-1}(p_2) \\ 
f_{r-1}(p_1) & f_{r-1}(p_0) &
f_{r-1}(p_2) \\ 
\end{pmatrix}\ ,\qquad f_r:= \frac {\psi_r} {\sqrt{\d z}}\ ,\ \ \wh f_r:= \frac {\wh \psi_r}{\sqrt{\d z}}
\ee
Then $\mathfrak V(z)$ solves a RHP with the jumps (\ref{nicejumps}), the asymptotic behavior 
\be
\mathfrak V(z) = {\rm diag} (C_1, C_2, C_3) (\1 + \mathcal O(z^{-1})) {\rm diag} (z^r, 1, z^{-r})
\ee
and Problem \ref{outerRHP} admits a solution if and only if $C_1C_2C_3\neq 0$.  
\ep
{\bf Proof.}
By Proposition \ref{propjumps} the jumps conditions are automatically fulfilled.  As $z\to \infty$, $p_j(z)\to \infty_j \in \L$ by construction.  Then 
\beas
&f_r(p_1)=  K_r^{(1)} z^r(1 + \mathcal O(z^{-1}))\ ,
f_r(p_0)=  K_r^{(0)} \frac 1 z (1 + \mathcal O(z^{-1}))\ , 
f_r(p_2)=  K^{(2)}_r z^{-r-1}(1 + \mathcal O(z^{-1})), \\
&\wh f_{r-1}(p_1)= \wh K_{r-1}^{(1)} z^{r-1}(1 + \mathcal O(z^{-1}))\ ,
\wh f_{r-1}(p_0)= \wh  K_{r-1}^{(0)}(1 + \mathcal O(z^{-1}))\ ,
\wh f_{r-1}(p_2)=  \wh K_{r-1}^{(2)}  z^{-r-1}(1 + \mathcal O(z^{-1}))\ .  
\eeas
This asymptotic behaviour follows from the divisor properties of $\psi_r, \wh \psi_r$; for example, 
 since  $\psi_r$ has a pole at most of degree $r+1$ at $\infty_1$ where $\sqrt{\d z}$ has a simple pole, $f_r$ must have at most a pole of order $r$ at infinity.

The last claim follows from the fact that the Problem \ref{outerRHP} requires that the constants $K_r^{(1)}, K_{r-1}^{(2)}$ and $\wh K_{r-1}^{(0)}$ should not vanish. It is only in this case that  the matrix $\mathfrak V$ can be normalized by a left multiplication to behave like $(\1 + \mathcal O(z^{-1})){\rm diag} (z^r,1,z^{-r})$.
To prove the necessity of the last claim we have to show that if any of the constants $K_r^{(1)}, K_{r-1}^{(2)}, \wh K_{r-1}^{(0)}$ vanish, then the Problem \ref{outerRHP} is unsolvable. To see this we first point out that standard arguments show that if a solution to Problem \ref{outerRHP} exists, then that solution is unique. Therefore the solution of Problem \ref{outerRHP} exists {\em if and only if} the only row-vector solution $\mathbf F(z)$ of the  RHP with the same jumps as \ref{outerRHP} but with the asymptotic condition (\ref{growthcond}) at $z=\infty$  replaced by
\be
\mathbf F(z) = [\mathcal O(z^{r-1}),  \mathcal O(z^{-1}) , \mathcal O(z^{-r-1})]
\ee
is the trivial solution $\mathbf F(z) \equiv 0$. If any of the constants above is zero, the corresponding row then provides a nontrivial solution precisely to this latter problem. This proves the necessity of the last claim.
{ \bf Q.E.D.}\par \vskip 5pt
\subsection{Theta--functional expressions}
\label{whyareyoureadingthesourcewhenyoucanreadthepdf}
The notation below is borrowed from \cite{BertolaAdvances} and the definitions can be found in \cite{Faybook} and  are reviewed in Appendix \ref{se:notation}.

The spinors $\psi_r, \wh \psi_r$ in Proposition \ref{propbehaviour} can be written ``explicitly'' in terms of Theta functions as follows
\bea
 \psi_r &\& = \frac {\Theta_\Delta(p-\infty_2)^{r} }{\Theta_\Delta(p-\infty_1)^{r+1}}
\Theta\le[\mathbf A \atop \mathbf B\ri] (p+ r\infty_2 -(r+1)\infty_1)h_\Delta(p)\\
\wh\psi_{r-1} &\&= \frac {\Theta_\Delta(p-\infty_2)^{r}}{\Theta_\Delta(p-\infty_1)^{r}\Theta_\Delta(p-\infty_0)}\Theta\le[\mathbf A \atop \mathbf B\ri] (p+ r\infty_2  - \infty_0 -r\infty_1)h_\Delta(p) .
\eea
Here  $h_\Delta$ 
is a certain fixed holomorphic spinor defined in (\ref{omegadelta}.
We now recall that $\Theta_\Delta(p-q)$, as a function of a point  $p$, has a simple zero at $p=q$ and at other $g-1$ points whose positions are independent of $q$ and depend solely on the choice of odd characteristic $\Delta$. Also, by construction \cite{Faybook}, the spinor $h_\Delta$ has zeroes precisely at the same $g-1$ points (and of the same multiplicity). It thus appears that $\psi_r$ has a zero of multiplicity $r$ at $\infty_2$ and  a pole of order $r+1$ at $\infty_1$. The order of the pole at $\infty_1$ may be smaller if the term $\mathcal T_r(p):= \Theta\le[\mathbf A \atop \mathbf B\ri] (p+ r\infty_2  -(r+1)\infty_1)$ vanishes there. If this happens, the constant $C_1$ in Proposition \ref{propbehaviour} would be zero and the RHP unsolvable. Similarly, if $\mathcal T_{r-1}(\infty_2)=0$ then $\psi_{r-1}$ has a zero of multiplicity {\em higher} than  $r$, and consequently $C_3=0$. Finally if $\wh{\mathcal T}_{r-1}(p):= \Theta\le[\mathbf A \atop \mathbf B\ri] (p+ r\inf
 ty_2  - \infty_0 - r\infty_1)$ vanishes at $\infty_0$, then the spinor $\wh \psi_{r-1}$ does not have a pole at $\infty_0$ and thus $C_2$ in Proposition \ref{propbehaviour} vanishes. Note that
$$
\mathcal T_r(\infty_1) = \mathcal T_{r-1} (\infty_2) = \wh {\mathcal T}_{r-1} (\infty_0)=\Theta\le[\mathbf A \atop \mathbf B\ri] (r\infty_2  - r\infty_1)\
$$
Therefore we have the proposition below
\bp
\label{taufunc}
The Riemann--Hilbert problem \ref{outerRHP} is solvable if and only if  
\be
\Theta\le[\mathbf A \atop \mathbf B\ri] (r\infty_2  - r\infty_1)\neq 0
\ee
\ep

We now need  to establish the nonvanishing property required by Proposition \ref{taufunc}. We recall that in our case $\mathbf A \in i\R^g$ and $\mathbf B=0$ (\ref{characters}). We then can use the results on bordered Riemann surfaces contained in \cite{Faybook}. 

\bt
\label{nonzerotau}
For any $ \mathbf A \in 2i\pi \R^g, \ r\in \Z$,
\be
\Theta\le[\mathbf A\atop 0\ri]( r\infty_2 - r\infty_1) \neq 0\ . \label{ifyoureadthissmile}
\ee
Therefore, Problem \ref{outerRHP} is always solvable.
\et
\br
The theorem could be rephrased as stating that the monodromy data for the outer parametrix do not lie on the Malgrange divisor. In this case the Malgrange divisor is actually identifiable with the $(\Theta)$ divisor in the Jacobian of the curve $\L$ \cite{Korotkin05}.
\er
{\bf Proof.}
Recall from Proposition \ref{Mcurve} that $\mathcal L$ has an antiholomorphic involution and can be represented as the {\em double of a bordered Riemann surface} (\cite{Faybook}, Chap. VI).

The $\beta$--cycles are homologous to cycles fixed by the antiholomorphic involution, since we can realize each of them as a path following the gap in $\R$ on one of the outer sheets and the same segment on the middle sheet.
There are in total $g+1$ cycles fixed by the involution; aside from the $g$  $\beta$--cycles, there is also  the cycle which we denote by $\beta_{g+1}$ that covers  the union of the segment $[b_0,a_0]$ on the middle sheet  with all the unbounded gaps in the three sheets (see Fig. \ref{rhosupp}). 

We now use Corollary 6.5 on page 114 of \cite{Faybook} which describes the intersection of the $\Theta$--divisor with a $g$--dimensional real  torus in the Jacobian of the form $\le[ i\R^g\atop \mu\ri]$ (with $\mu=(\mu_j)\in \frac 12 \Z^g$ a half--period). It consists of the image in the Jacobian of the set of divisors of degree zero that have a form $ \DD_{g-1}  - \DD^0_{g-1}$, where
$ \DD^0_{g-1}$ is the divisor of degree $g-1$, such that $ 2\DD^0_{g-1}$ is linearly equivalent to the canonical divisor and $ \DD_{g-1}$ is any positive  divisor of degree $g-1$ invariant under the involution and such that $\DD_{g-1}$ has $(1+2\mu_j)\mod 2$ points in $\beta_j$ for $j=1,\ldots,g$.
This means that if there existed a point $\mathbf A\in  i\R^g$ such that  $\Theta\le[\mathbf A\atop 0 \ri](0)=0$, then it would be the image of a positive divisor of degree $g-1$ with exactly one point  in each of the $g$ gaps, a clear contradiction (see also Proposition 6.16 in \cite{Faybook}). This proves that $\Theta\le[\mathbf A\atop 0 \ri](0)$ cannot vanish for any value of $\mathbf A\in i\R^g$.

To complete the proof, recall that 
$\Theta\le[\mathbf A \atop 0 \ri] (r\infty_2  - r\infty_1)$ is proportional to $ \Theta(\mathbf A + r\infty_2  - r\infty_1)\ .$
 To prove that the latter expression is nonzero  we only need to show that
 $\mathfrak u(r\infty_2  - r\infty_1) \in i \R^g$, where $\mathfrak u$ is the Abel map. By our choice of $\alpha$--cycles and  normalization \eqref{normal} , holomorphic  differentials $\omega_j$ satisfy $\bar\omega_j(\bar z)=-\omega_j(z)$.
 Therefore an image under the Abel map of any divisor of degree $0$ contained in $\beta_{g+1}$ is purely imaginary. Since $\infty_{1,2}\in \beta_{g+1}$, the proof is complete. \QED

The expressions in terms of Theta functions for the columns of the outer parametrix $\Psi$ possess other interesting relations that were investigated in \cite{Bertola:EffectiveIMRN}. 
\subsubsection{Genus $0$ case}
\label{genus0}
A particularly simple situation is the case in which the surface $\L$ is of genus $0$, namely there are only two cuts, one in $\R_+$ and one in $\R_-$. In this case we can write quite explicit algebraic expressions for all the objects above (see Fig. \ref{figthreeplanes} for an example). First, 
we introduce a uniformizing parameter $t:\ \C P^1\to \L$ in terms of which the meromorphic function $z:\L \to \C P^1$ is written as 
\be
z:= u_0  t + C+ \frac {u_1}{t-1} + \frac {u_2}{t+1}\label{Xunif}.
\ee
Here we have identified the three poles $\infty_{0,1,2}$ with $t=\infty,1, -1$ respectively. 
Once the uniformization has been fixed, the expressions for the spinors above are extremely simple
\be
\begin{split}
& f_r:= \frac {\psi_r}{\sqrt{\d z}} = \frac {(t+1)^r}{ (t-1)^{r+1} }\sqrt {\frac{\d t}{\d z}}\ ,\qquad f_r =  
\frac {(t+1)^{r+1}}{(t-1)^{r} \sqrt {
u_0 (t^2-1)^2 - u_1 (t+1)^2 - u_2(t-1)^2\ ,
}}
\\
& 
\wh f_r := \frac {\wh \psi_r}{\sqrt{\d z}} = \frac{(t+1)^{r+1}}{(t-1)^{r+1}} \sqrt{\frac {\d t}{\d z}}
\ ,\qquad
\wh f_r = \frac { (t+1)^{r+2}}  { (t-1)^r \sqrt{u_0 (t^2-1)^2 - u_1 (t+1)^2 - u_2(t-1)^2}} \ .
\end{split}
\ee
The normalization is obtained by expanding near the three points at infinity
\bea
f_r\sim \le\{
\begin{array}{cc}
\ds z^r \frac{2^r}{(-u_1)^{r+\frac 1 2 }} & t\sim 1\\[5pt]
\ds\frac {i}{z^{r+1}} \frac {u_2^{r+\frac 1 2}}{2^{r+1}} & t\sim -1\\[5pt]
\ds \frac{\sqrt{u_0}}{z} & t\sim 0
\end{array}
\ri.
\ ,\ \ \ 
\wh f_{r-1}\sim \le\{
\begin{array}{cc}
\ds z^{r-1} \frac {2^{r}}{(-u_1)^{r-\frac 1 2}} & t\sim 1\\[5pt]
\ds
\frac {-i} {z^{r+1}} \frac {u_2^{r+\frac 1 2}}{2^{r}}
 & t\sim -1\\[5pt]
\ds \frac 1 {\sqrt{u_0}}
& t\sim 0 
\end{array}\ .
\ri.
\eea

The solution to the RHP problem Problem \ref{outerRHP} is
\be
\label{paramzero}
\Psi(z) ={\rm diag}( (-u_1)^{r-\frac 1 2} 2^{-r}, i \sqrt {u_0} ,i u_2^{-r+\frac 1 2})\begin{pmatrix} 
f_r(t_1(z))&
 f_r(t_0(z)) &
 f_r(t_2(z))\\
\wh f_{r-1}(t_1(z))& 
\wh f_{r-1}(t_0(z)) & 
\wh f_{r-1}(t_2(z))\\
f_{r-1}(t_1(z))& 
 f_{r-1}(t_0(z)) & 
 f_{r-1}(t_2(z))
\end{pmatrix}\ .
\ee
Here $t_{1,2,3}(z)$ are $t$-coordinates of sections $p_{1,2,3}$ discussed in Section \ref{DomainD}.
\subsection{Local parametrix (solution of Problem \ref{localRHP})}
\label{SectionAiry}
For the sake of completeness we spell out the form of the parametrix near a endpoint of a cut  where the density $\rho_{1,2}$  vanishes like a  square root (see Corollary \ref{squareroot}).

\subsubsection{The rank-two parametrix}

\begin{wrapfigure}{r}{0.6\textwidth}
\resizebox{0.3\textwidth}{!}{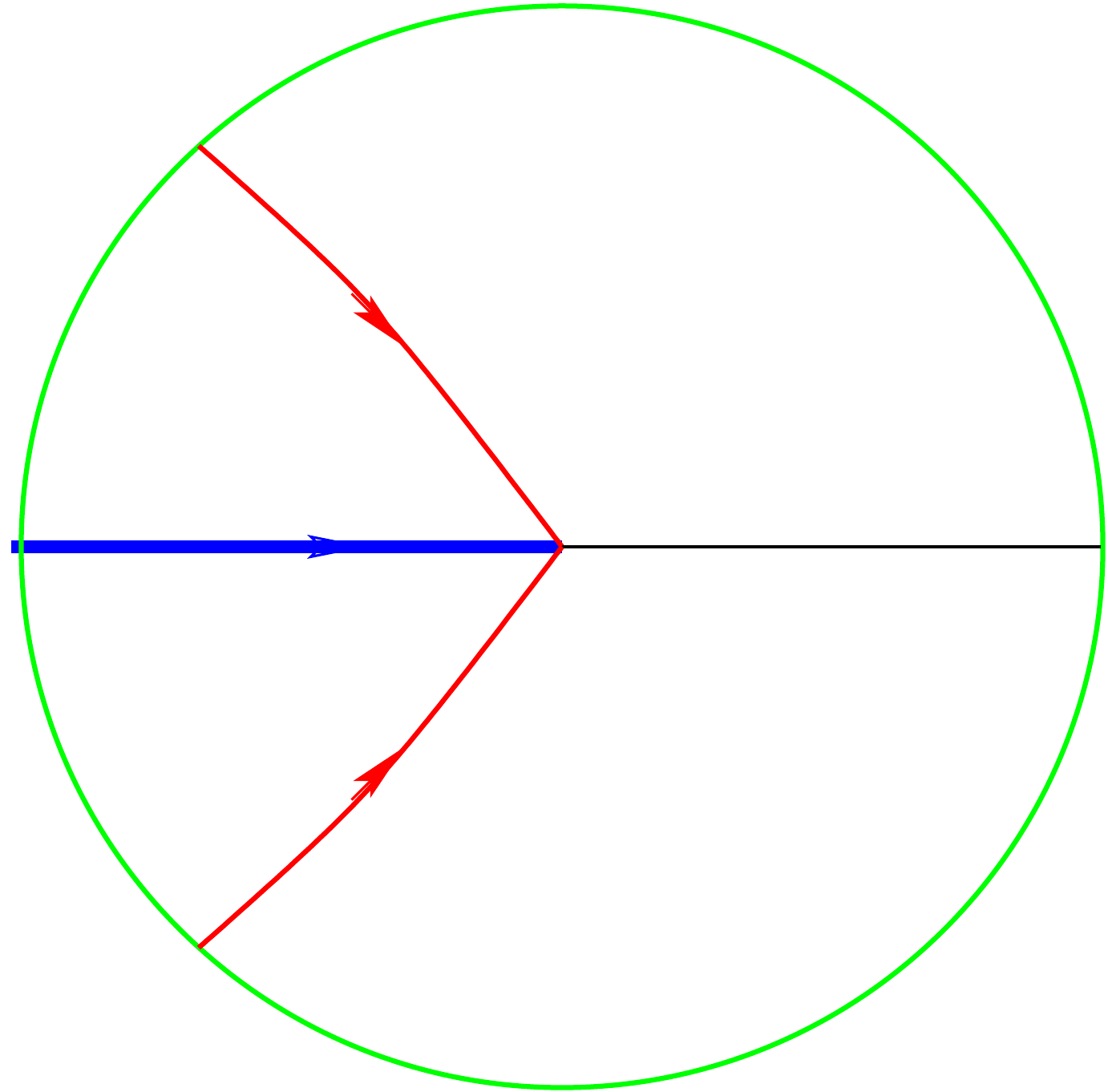}\resizebox{0.3\textwidth}{!}{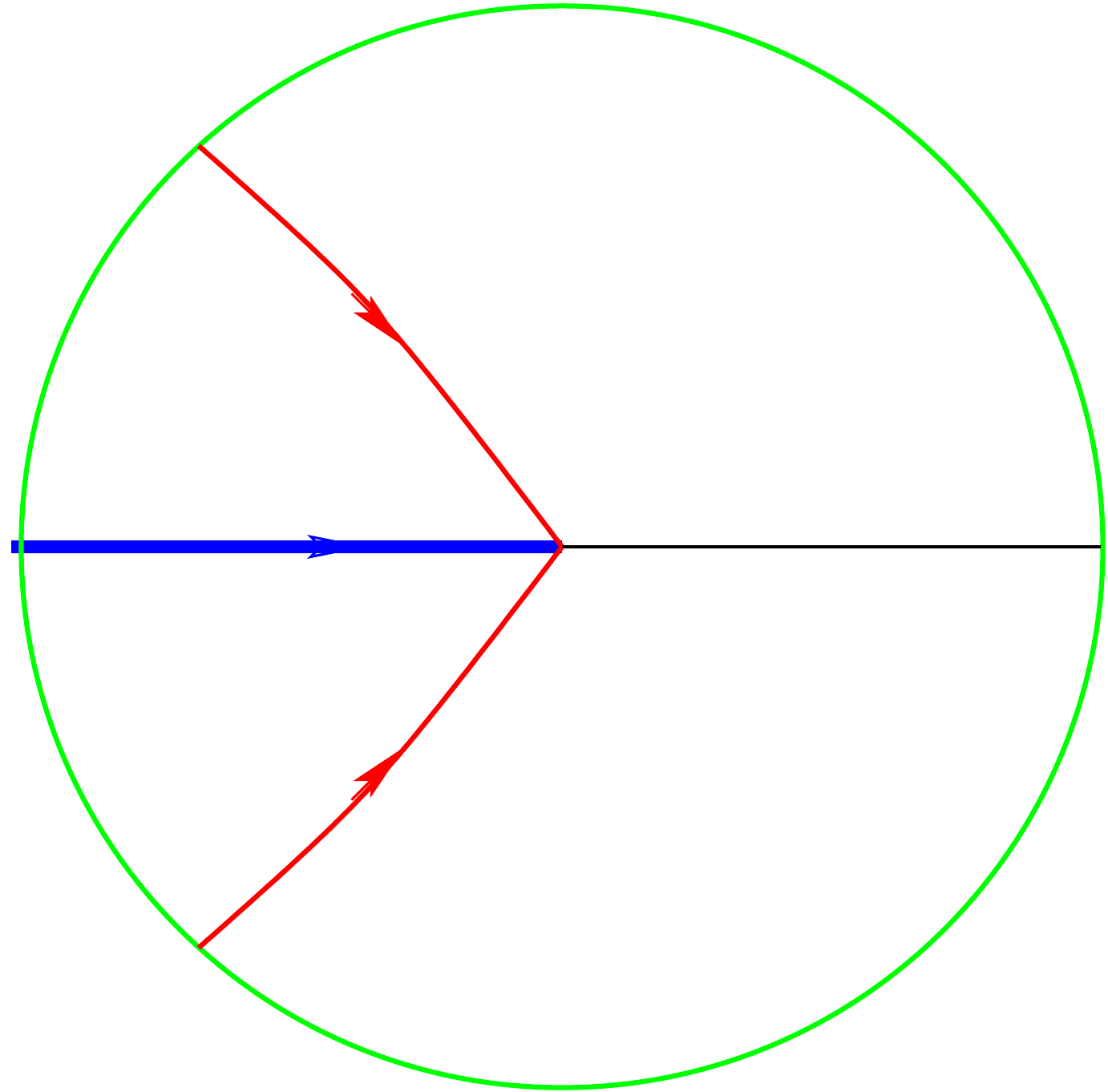}
\caption{The exact Riemann--Hilbert problem near a right endpoint in $\R_+$, and in terms of the zooming local coordinate $\xi$.}
\label{Airyparam}
\end{wrapfigure}

We need to solve the exact RHP in Fig. \ref{Airyparam}.
This part is essentially identical to the established results in \cite{Deift, DKMVZ}.
We consider only the previously defined neighborhood $\mathbb D_a$ of the right endpoint $a:=a_{2\ell-1}$ of one of the intervals of $\rho_1$. The modifications for the other cases are straightforward. 
Near $x=a$ the effective potential  $\varphi_{_1}$ is a piecewise analytic function with the jump $
\varphi_{_1+}-\varphi_{_1-} = 4i\pi\epsilon_\ell$, 
where $\ell$ is the number of the gap that starts on the right of $a$ and $\epsilon_\ell$ is the cumulative filling fraction (\ref{cumulativeepsilon}).

Let $\varphi$ be a function defined by
$$\varphi(z)=\left \{\begin{array}{cc} \varphi_1(z) +2 i\pi(1-\epsilon_\ell)\ , & \Im z>0\\
\varphi_1(z) + 2 i \pi(1+ \epsilon_\ell)\ , & \Im z\leq 0
\end{array}
 \right .\ .$$
Then $\varphi$ has an analytic expansion in Puiseux series of the form described below.

\bl
\label{lemmaPuiseux}
The locally analytic function $\varphi$ has an expansion 
\bea
\varphi (x) = \frac 43 C( x-a)^{\frac 32}(1 + \mathcal O(x-a))\cr
 C =  \lim_{x\to a^-} \frac {\pi \rho_1(x)}{\sqrt{a-x}}>0 \nonumber`
\eea
where the cut of the root extends on  the left of $a$ and the  term $\mathcal O(x-a)$ is analytic at $a$ (i.e. has a convergent Taylor series).
\el
{\bf Proof.}
Recalling the representation (\ref{Yrep}) for the $\gg$ functions we have
\be
\varphi(z) = \int_a^z \frac {\d \varphi_{_1}(\xi)}{\d \xi}\d \xi = \int_a^z\le( Y^{(1)}(\xi)- Y^{(0)}(\xi)\ri) \d \xi\ .
\ee
Now a small loop around $a$ interchanges $Y^{(1)} \leftrightarrow Y^{(0)}$, which means that 
\be
 Y^{(1)}(\xi)- Y^{(0)}(\xi) = \sqrt{\xi-a}\, (2C + \mathcal O(\xi-a))\label{54}
\ee
with the term $\mathcal O(\xi-a)$ analytic.
The expression for $C$ is obtained by recalling that by Def. \ref{effdef}
\be
 \varphi'(z)=\varphi_{_1}'(z) =  V_1'(z) - 2\int_{0}^{+\infty} \frac {\rho_1(s)\d s}{z-s} - \int^{-\infty}_{0} \frac {\rho_2(s)\d s}{z-s}
\ee
 Now $\rho_1(s) \sim C\sqrt{a-s}$ near $s=a$ (with $C>0$) due to (\ref{54}) and $\varphi'_+ - \varphi_-' = 4\pi i \rho_1$.
 The sign of $C>0$ is due to the fact 
 that $\Re \varphi_{_1} = \Re \varphi$ and it must be positive in the gap ($x>a$). 
\QED
\br
In a non-generic situation, one can only conclude that 
\be
\varphi(z) = (z-a)^{k+\frac 32} (\wt C + \mathcal O(z-a))\ ,\ \ \wt C > 0
\ee
for some nonnegative integer $k$. This case is called {\em nonregular} and it is treated in \cite{DKMVZ}. The construction there applies identically in this case, but we prefer to concentrate on a detailed discussion  for the most generic case.
\er

We define the {\bf zooming local parameter} by the equation
\be
\label{zoom}
\frac 43 \xi^{\frac 32} = N \varphi(z)
\ee
so that the fixed-size neighborhood of  $a$ in the $z$--plane is mapped conformally (one-to-one) to an homothetically expanding neighborhood of the origin in the $\xi$ plane with a diameter growing like $N^{\frac 23}$. 
The choice of the root in \eqref{zoom} is such that the cut is mapped to $\R_-$ of $\xi$-plane.

We then introduce the standard rank--two Airy parametrix $\mathcal R^0(\xi)$
\cite{Deift, DKMVZ} as the piecewise defined matrix  $\mathcal R^0_j$
 constructed in terms of the Airy function $\mbox{Ai}(x)$ as follows. If $\boldsymbol{\sigma}_3=\left (\begin{array}{cc} 1 & 0\\ 0 & -1\end{array} \right )$, define
\bea
\mathcal R^0_k(\xi)&\&:= \sqrt{2\pi} {\rm e}^{-\frac {i \pi}4}
 \le\{
\begin{array}{llc }
\begin{pmatrix} y_0 & -y_2\cr
y_0'& - y_2' \end{pmatrix} {\rm e}^{\le(\frac 23\xi^{\frac 32} + i\pi N \epsilon_{_\ell}\ri)\boldsymbol{\sigma}_3}
=&
\begin{pmatrix} y_0 & -y_2\cr
y_0'& - y_2' \end{pmatrix} {\rm e}^{\frac N2\varphi_{_1} \boldsymbol{\sigma}_3}
 & k=1\\[15pt]
\begin{pmatrix} -y_1 & -y_2\cr
-y_1'& - y_2 \end{pmatrix}
{\rm e}^{\le(\frac 23\xi^{\frac 32}+i\pi N \epsilon_{_\ell}\ri)\boldsymbol{\sigma}_3}=&
\begin{pmatrix} -y_1 & -y_2\cr
-y_1'& - y_2 \end{pmatrix}
{\rm e}^{\frac N2 \varphi_{_1}\boldsymbol{\sigma}_3} & k=2\\[15pt]
\begin{pmatrix}-y_2 & y_1\cr
-y_2'&  y_1'\end{pmatrix} {\rm e}^{\le(\frac 23\xi^{\frac 32}-i\pi N \epsilon_{_\ell}\ri) \boldsymbol{\sigma}_3}=
&
\begin{pmatrix}-y_2 & y_1\cr
-y_2'&  y_1'\end{pmatrix} {\rm e}^{\frac N2 \varphi_{_1}  \boldsymbol{\sigma}_3}=
 & k=3\\[15pt]
\begin{pmatrix} y_0 & y_1\cr
y_0'& y_1' \end{pmatrix} {\rm e}^{ \le(\frac 23\xi^{\frac 32}-i\pi N \epsilon_{_\ell}\ri)  \boldsymbol{\sigma}_3}=
&
\begin{pmatrix} y_0 & y_1\cr
y_0'& y_1' \end{pmatrix} {\rm e}^{ \frac N2 \varphi_{_1} \boldsymbol{\sigma}_3} & k=4
\end{array}
\ri.,
\eea
where index $k$ refers to regions in Fig. \ref{Airyparam} and, for $s=0,1,2$,
\be y_s:= \omega^s {\rm Ai}(\omega^s \xi),\ \ \ \ \omega = {\rm e}^{2i\pi/3}.\ee
Each $\mathcal R_k^{0}$ has the following uniform asymptotic behavior near $\xi=\infty$.
\be
\mathcal R_k^{0}(\xi) =
\xi ^{-\frac {\boldsymbol{\sigma}_3}4}\frac 1 {\sqrt{2}} \begin{pmatrix} 1 &  1\cr -1 & 1\end{pmatrix} {\rm e}^{\le(-\frac{i\pi}4 \pm i\pi N \epsilon_{_\ell}\ri) {\boldsymbol{\sigma}_3} }
({\bf 1} +\mathcal O(\xi^{-3/2})).\label{formal}
\ee
where the $\pm$ depends on the half-plane (upper/lower) in which the asymptotics is considered.

Thus, the final form of the local parametrix is defined as a matrix $\mathcal R$ whose restrictions to the four regions in Fig. \ref{Airyparam} are
\be
\mathcal R_k(\xi):=
\overbrace{{\rm e}^{-\le(-\frac{i\pi}4 \pm i\pi N \epsilon_{_\ell}\ri)\boldsymbol{\sigma}_3 }\frac 1{\sqrt {2}} \le[
\begin{array}{cc}
1 &-1\\
1&1
\end{array}
\ri] \xi^{\frac {\boldsymbol{\sigma}_3} 4}}^{:=F(z)}
\mathcal R_k^{0}(\xi)\ .
\ee
Here
\begin{enumerate}
\item[(i)] the prefactor $F(z)$ solves a RHP  on the left
\be
\begin{split}
F(z)_+ & =\le[
\begin{array}{cc}
0 & -1\\
1 & 0
\end{array}
\ri] F(z)_-\ ,\qquad \xi(z)\in \R_-\ \\
F(z) _+ & =  {\rm e}^{-2i\pi N\epsilon_{_\ell}{\sigma}_3}F(z)_-\ ,\qquad \xi(z)\in \R_+\ .
\end{split}
\ee
\item[(ii)]  $\mathcal R$ satisfies the exact jump conditions  of the RHP in Fig. \ref{Airyparam} , except for the jump condition on the cut :
\be
\begin{split}
\mathcal R_+ & =\le[
\begin{array}{cc}
0 & -1\\
1 & 0
\end{array}
\ri] \mathcal R_- \le[
\begin{array}{cc}
0 & 1\\
-1 & 0
\end{array}
\ri]  \ \ \ \  \xi \in \R_-\\
\mathcal R_+ & =
{\rm e}^{-2i\pi N \epsilon_{_\ell} \boldsymbol{\sigma}_3}  \mathcal R_- {\rm e}^{2i\pi N \epsilon_{_\ell} \boldsymbol{\sigma}_3} \ \ \ \ \  \xi\in \R_+
\end{split}.
\ee 
\item[(iii)]   $\mathcal R(\xi) = \1 + \mathcal O(N^{-1})$ uniformly on the boundary of the neighborhood $\mathbb D_a$.
\end{enumerate}
\subsection{Rank-three parametrix}
Now we embed the rank-two parametrix from the previous section into a matrix 
\be
{\mathcal R_a}(x) = \mathcal R(x) \oplus 1 =: \le[
\begin{array}{c|c}
\mathcal R(x) & \\
\hline 
&1
\end{array}
\ri] \ .
\label{cherotturadicazzi}
\ee
Let define the parametrix $\mathcal P_a(z)$ within the disk $\mathbb D_a$ as 
\be
\mathcal P_a(z) := \Psi(z) \mathcal R_a(z)\ ,
\ee
where $\Psi$ is the outer parametrix constructed earlier.
The only point to address now  is whether  $\Psi (z) \mathcal R_a(z)$ is {\em bounded} inside the disk, since near $a$
\be
\Psi(z) = \mathcal O( (z-a)^{-\frac 14}),\quad F(z)= \mathcal O( (z-a)^{-\frac 14}).
\ee
Thus the product $\Psi (z) (F(z)\oplus 1)$ may at most have square root singularities: however, comparing the RHPs that $\Psi$ and $F$ solve, we see that the product is a single-valued matrix, thus it must be analytic since at worst it may have singularities of type $(z-a)^{-\frac 12}$. Being single--valued, however, it must have a Laurent series expansion that must , in fact, be a Taylor series by the previous a priori estimate of its growth. Therefore the product is actually analytic. Since $F(z)$ is solely responsible for the unboundedness of $\mathcal R$,  this proves that $\mathcal P_a(z)$ is bounded.

\subsection {Asymptotics of the biorthogonal polynomials}
Let us go back to modifications discussed in Section 4.1 of the RHP Problem \ref{RHPp} for Cauchy biorthogonal polynomials. It follows from \eqref{maintransform}, that
within the  half-lenses depicted in Fig. \ref{FigGamma0}, \ref{FigGamma0left}  (on the $\pm $ sides of the cuts) the original matrix $\Gamma(z)$ reads
\bea
\Gamma_\pm (x) =C_\gamma \ZZ (x)_\pm G_\pm^{-1} C_\gamma^{-1} \le[
\begin{array}{ccc}
1 &  & \\
 \pm {\rm e}^{N V_1}&1&\\
 &&1
\end{array}
\ri]\ ,\ x\in \mathcal A\\
\Gamma_\pm(x) =C_\gamma \ZZ (x)_\pm G_\pm ^{-1} C^{-1}_\gamma \le[
\begin{array}{ccc}
1 &  &\\
&1& \\
 &\pm {\rm e}^{N V_2}&1
\end{array}
\ri]\ ,\ x\in \mathcal B.
\eea

According to \eqref{gammamain}, the  degree $n$ monic biorthogonal polynomial  is 
\be
\begin{split}
p_n(x) &= \Gamma_{11}(x) = {\rm e} ^{\frac N2\le(V_1 + \gg^{(0)}_{\pm} + \gg^{(1)}_{\pm} + \gamma_+ \ri)} \le[\ZZ_{11\pm} {\rm e}^{-\frac  N2 \varphi_{_1\pm } } \pm \ZZ_{12\pm} {\rm e}^{\frac  N2 \varphi_{_1\pm} }\ri] = \\
&=  {\rm e} ^{\frac N2\le(V_1 - \gg^{(2)}  + \gamma_+ \ri)} \le[\ZZ_{11\pm} {\rm e}^{-\frac  N2 \varphi_{_1\pm } } \pm \ZZ_{12\pm} {\rm e}^{\frac  N2 \varphi_{_1\pm} }\ri]
\end{split}
\ \ \ \ \ (x\in \mathcal A).
\ee
Note that for $x>0$, $\gg^{(2)}(x) \in \R$  and has no jumps,  and for $x\in \mathcal A$ we have $\varphi_{_1\pm} = -2i \Im\gg^{(1)}_\pm (x) = \pm 2i\pi \left ( \int_{a_0}^x \rho_1(s)\d s-1\right )$. From (4-2)--(4-9), 
\be
\ZZ(z) = \mbox{diag}(1,-1,1)\ov {\ZZ(\ov z)}\mbox{diag}(1,-1,1)\  (z\in \C)\ ,\ \ \ZZ_{12\pm}(x) = \pm \ZZ_{11\mp}(x)= \pm \bar{\ZZ}_{11\pm}(x)\ (x\in \mathcal A),
\ee
and thus 
\be
p_n(x) =  {\rm e} ^{\frac N2\le(V_1 - \gg^{(2)}  + \gamma_+ \ri)}2 \Re \le[\ZZ_{11} {\rm e}^{-i N\pi \int_{a_0}^{x} \rho_1(s)\d s }\ri]\ ,\ \ \ \ \ \ x\in \mathcal A.
\ee
Vice versa, for $z\not\in \mathcal A$ (and outside of the right lenses) one has 
\be
p_n(z) = {\rm e}^{N \gg^{(1)}} \ZZ_{11}(z)
\ee
Recall that $\ZZ = \le(\1 + \mathcal O(N^{-1})\ri) \wh \ZZ$ , where $\wh \ZZ$ is defined in \eqref {whGamma}. Then  one can obtain  uniform asymptotic information on the behaviour of $p_n$ in any compact set of the complex plane. In particular, away from the endpoints, the asymptotics is expressible in terms of Theta functions (genus $\geq 1$) or algebraic expressions (genus $0$).

To obtain asymptotic information for the biorthogonal companions $q_n$, one simply interchanges the r\^oles of the measures.

\section{Asymptotic spectral statistics and universality}
\label{universality}
Once we have obtained a uniform asymptotic control of the biorthogonal polynomials, it is natural to investigate, in parallel to what has been done for the Hermitean matrix model (see, e.g. \cite{DKMVZ}), the large $N$ behavior of the correlation functions for the Cauchy two-matrix model studied in \cite{Bertola:CauchyMM} .
We recall that the finite--size correlation functions for the spectra of $M_1, M_2$ distributed according to \eqref{2matrix} correspond to a multi--level determinantal point process with the kernel
\bea
\mathbb{K}_N(x,y)&\&:=  
\sum_{j=0}^{N-1} \frac {p_j(x)q_j(y)}{h_j}\ ,\\
&\& \iint_{\R_+^2} p_j(x)q_i(y) \frac { {\rm e}^{-N(V_1(x) + V_2(-y))}}{x+y} = h_j\delta_{ij} \ ,
\eea
where $\{p_j(x), q_j(y)\}_{j\in \N}$ are the {\em monic} biorthogonal polynomials.\footnote{Normalizing factors $h_j$ appear in the definition of $\mathbb{K}_N(x,y)$ in contrast to the formula (3.31) in \cite{Bertola:CauchyMM} since here we use monic rather than orthonormal biorthogonal polynomials.}

Define four auxiliary kernels 
\bea
&& H_{00}(x,y):= \mathbb{K}_N(x,y)\ , \\
&& H_{10}(y',y):= \int_{\R_+} \frac{{\rm e}^{-N V_1(x)}\d x}{y'+x} H_{00}(x,y)\ ,\\
&& H_{01} (x,x'):=\int_{\R_+} \frac{{\rm e}^{-N V_2(-y)}\d y}{y+x'} H_{00}(x,y)\ ,\\
&& H_{11} (y,x):=
 \int_{\R_+}\int_{\R_+} H_{00}(x',y')\frac{{\rm e}^{-N (V_1(x')+ V_2(-y'))} \d x' \d y'}{(y+x')(x+y')} - \frac 1{x+y}\ .
\eea

\br To obtain kernels for the correlation functions auxiliary kernels need to be multiplied by appropriate exponentials involving the potentials $V_1,V_2$.

\er

\bp {\rm{(Proposition 3.2 in \cite{Bertola:CauchyMM})}}
\label{prop51}
\footnote{\label{foot5} We adjusted the formulae from \cite{Bertola:CauchyMM}, see footnote \ref{foot6}.}
The auxiliary kernels  are given in terms of the solution of the RHP in Proposition \ref{RHPp} as follows (for $x,x',y,y'\geq 0$)
\be
\begin{array}{cll}
H_{00}(x,y)&\&\ds = -\frac 1{(2i\pi)^2} \frac {\le[  \Gamma^{-1}(-y)  \Gamma(x)\ri]_{3,1} }{x+y}  \ ,\qquad 
H_{10}(y',y) \ds =\frac 1{2i\pi} \frac {\le[  \Gamma^{-1}(-y)  \Gamma(-y')\ri]_{3,2} }{y'-y}\ ,  \\[10pt]
H_{01}(x,x') &\& \ds = \frac 1{2i\pi} \frac {\le[ \Gamma ^{-1}(x')  \Gamma (x)\ri]_{2,1} }{x- x'}\ ,\qquad
H_{11}(y,x)  \ds = \frac {\le[ \Gamma ^{-1}(x)  \Gamma(-y)\ri]_{2,2} }{x+y} \ .
\end{array}
\label{Kernels}
\ee
\ep
One should not expect  that the kernels $H_{00}(x,y),  H_{11}(y,x)$, $x,y>0$  display any universal behavior, even in the scaling regime
$
x = \frac \xi {N^\alpha} \ ,\  y = \frac {\eta}{N^\alpha}
$.
The only ``interaction'' between the two matrices $M_1$ and $M_2$ that can lead to a universality class is near the zero eigenvalue.

On the other hand, if one considers separately statistics of the eigenvalues of $M_1$ or $M_2$, no new universality phenomena will appear, as we briefly explain below. It is sufficient to consider $H_{01}(x,x')$ since the computation for  $H_{10}(y,y')$ is completely analogous. 
Again, we consider only the regular case. 
 
 \paragraph{Universality in the bulk.}
Let $c$ belong to the interior of some cut. For simplicity we will consider only $\mathcal A$, but the result extends to the other case with obvious modifications in the roles of the potentials. Due to regularity assumptions, at $c$ we have $\rho_1(c) = C>0$. Consider 
\be
x = c + \frac {\xi}{C N}\ ,\qquad x' = c+ \frac {\eta}{CN}.
\ee
Then a straightforward computation using Proposition \ref{prop51} yields
\bea
&&\frac 1{ \rho_1(c)  N} H_{01}(x,x') {\rm e}^{-\frac N2 (V_1(x) +V_1(x'))} =\qquad \nonumber
\\  
&&\qquad \frac { {\rm e}^{-\frac N 2\le( \gg^{(2)}(x)-\gg^{(2)}(x')\ri)}}{2\pi i (\xi-\eta)} \le(
 {\rm e}^{\frac N2 \le( \varphi_{_1+}(x) - \varphi_{_1+}(x') \ri)} -  {\rm e}^{-\frac N2 \le( \varphi_{_1+}(x) - \varphi_{_1+}(x') \ri)}
 \ri) (1+ \mathcal O(N^{-1})) \ .
\eea
Recall that 
\be
\varphi_{_1+}(x) =  i \Im \varphi_{_1+}(x) = -2i\pi \int^{\infty}_x \rho_1(s)\d s
\ee
 so that 
 \be
\lim_{N\to\infty} \frac 1{\rho_1(c)  N} H_{01}(x,x') {\rm e}^{-\frac N2 (V_1(x) +V_1(x'))}  ={\rm e}^{-\frac {{ \gg^{(2)}}'(c)}{ 2 \rho_1(c)} (\xi -\eta)}\underbrace{\frac{ \sin (\pi(\xi-\eta))}{\pi (\xi-\eta)}}_{K_{\sin}(\xi,\eta)} =: {\wh K}_{\sin} (\xi,\eta)
 \ee
The prefactor above to the usual sine-kernel $K_{\sin}(\xi,\eta)$ is not universal (it depends on the densities), however since $\wh K_{\sin}$ is conjugate to $K_{\sin}$,
\be
\det \wh K_{\sin}(\xi_i, \xi_j) = \det K_{\sin} (\xi_i, \xi_j)\ .
\ee
Therefore all spectral statistics, gap probabilities etc. for $M_1$ or $M_2$ separately will follow the standard universality results for the Hermitean matrix model.

\paragraph{Universality at the edge.}
Similarly, one finds near the edge $x=a$ 
\bea
&& x =  a + \frac {\xi}{C N^{\frac 23}}\ ,\qquad 
 x' = a + \frac {\eta}{C N^{\frac 23}}\ ,\qquad  C:= \lim_{s\to a^-} \frac {\pi \rho_1(s)}{\sqrt{a-s}}\ ,\\
&\&\lim_{N\to \infty} \frac 1{C N^{\frac 23}} H_{01}(x,x') {\rm e}^{-\frac N2 (V_1(x) +V_1(x'))}  = \frac {Ai(\eta) Ai'(\xi) - Ai'(\eta) Ai(\xi)}{\xi - \eta} \ .
\eea

\appendix

\renewcommand{\theequation}{\Alph{section}.\arabic{equation}}

\newpage
\section{An example: ``double Laguerre'' biorthogonal polynomials}
\label{DoubleJac}
\begin{wrapfigure}{r}{0.5\textwidth}
\vspace{-20pt}
\resizebox{0.5\textwidth}{!}{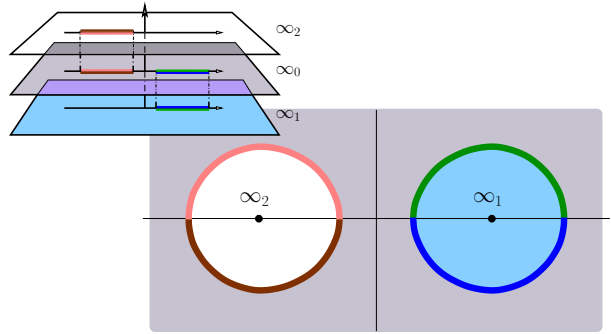}
\caption{Depiction of the three sheets and how they are mapped onto three disjoint regions of the $t$--plane. The real $x$--axis is mapped to the real $t$-axis and the two sides of each cut are mapped to the boundaries of the oval-shaped regions. The intersection of the ovals with the real $t$--axis are the four ramification points of the map $z(t)$.\label{figthreeplanes}}
\vspace{10pt}
\end{wrapfigure}

We consider the case where the two potentials are the same  $V_1(x)=V_2(-x)$ and are of the simplest possible form
\be
{\rm e}^{-V_1(x)} = x^a{\rm e}^{-bx}\ ,
\ee
where both $a,b>0$.

We can rescale the axis  and set $b=1$ without loss of generality.

The curve \eqref{algcurve} appearing in Theorem \ref{speccurve} was computed in \cite{BertolaBalogh, Bertola:CauchyMM} :
\be
y^3 - \le(\frac 1 3 + \frac {a^2}{z^2} \ri)y -\le( \frac {2a^2+6a+1}{3z^2} - \frac 2 {27} \ri)=0.
\ee
 According to section \ref{genus0} we find a  rational uniformization of this curve as 
 \bea
 z &\&= (1+a) t + \frac {2a+1}{2a+2}\le( \frac 1{t-1} +  \frac 1{t+1}\ri)\ ,
 \\
y&\& =
\frac {2a+1}{2a} \le( \frac 1{(a+1)t-a} - \frac 1{(a+1)t+a}\ri) - \frac 23.
 \eea
For $a>0$ there are four symmetric branch points on the real axis and the inner ones tend to zero as $a\to 0$, whereas all four tend to infinity as $\pm (a \pm 2 \sqrt{a}) + \mathcal O(1)$ as $a\to\infty$. Explicit formul\ae\ for the parametrix 
can be obtained by substituting $u_0=1+a,\ u_1=u_2=\frac{2a+1}{2a+2}$ into \eqref{paramzero}.

\section{Notation and main tools}

\label{se:notation}
For a given smooth compact  curve $\mathcal L$ of genus  $g$ with a fixed choice of symplectic homology basis  of $\alpha$ and $\beta$-cycles,  we denote by
$\omega_\ell$ the normalized basis of holomorphic differentials
\be
\label{normal}
\oint_{\alpha_j} \omega_\ell  = \delta_{j\ell}\ ,\qquad \oint_{\beta_j} \omega_{\ell} = \tau_{j\ell} = \tau_{\ell j}\ .
\ee
We will denote by
$\Theta$ the Theta function
\be
\Theta (\mathbf z):= \sum_{\vec n\in \Z^g} {\rm e}^{i\pi \vec n\cdot \tau \vec n - 2i\pi \mathbf z\cdot \vec n}\ . 
\ee
The Abel map $\mathfrak u: \mathcal L\to \C^g$ with  a base-point $p_0$ is
\be
\mathfrak u(p) = \le[\int_{p_0}^p \omega_1,\dots, \int_{p_0}^p \omega_g\ri]^t
\ee
and is defined up to the {\em period lattice} $\Z + \tau \cdot \Z$.
For brevity we will omit any symbolic reference to the Abel map when it appears as argument of a Theta function: namely if $p\in \mathcal L$ is a point and it appears as an argument of a Theta-function, the Abel map will be implied, meaning that
$$
\Theta(p-q) \hbox{ stands for } \Theta(\mathfrak u(p ) - \mathfrak u(q))\ .
$$
We denote by $\mathcal K$ the vector of Riemann constants 
\be
\mathcal K_j = -\sum_{\ell=1}^{g} \le[ \oint_{a_\ell} \omega_\ell(p) \int_{p_0}^p \omega_j(q) - \delta_{j\ell} \frac
{\tau_{jj}}2   \ri]\ ,
\ee
where the cycles $\alpha_j$ are realized as loops with basepoint $p_0$ and the inner integration is done along a path lying in the canonical dissection of the surface along the chosen representatives of the basis in the homology of the curve.

The Riemann constants have the crucial property that for a nonspecial divisor $\mathcal D$ of degree $g$, $\mathcal D = \sum_{j=1}^g p_j$, then the function
\be
f(p) = \Theta(p - \mathcal D - \mathcal K)
\ee
has zeroes precisely and only at $p=p_j$, $j=1\dots g$.

We also use Theta functions with (complex) characteristics: for any two complex vectors $\vec \epsilon, \vec \delta$ the Theta function with half-characteristics $\vec \epsilon, \vec \delta$ is defined via
\bea
\Theta\le[{\vec \epsilon \atop \vec \delta}\ri] (\mathbf z) := \exp\le(2i\pi\le( \frac {\vec \epsilon \cdot \tau \cdot \vec \epsilon }8 + \frac 1 2 \vec \epsilon \cdot \mathbf z + \frac 1 4 \vec \epsilon \cdot \vec \delta\ri) \ri) \Theta\le(\mathbf z + \frac {\vec \delta} 2 + \tau \frac {\vec \epsilon }2 \ri) \ .
\eea
Here the half-characteristics of a point are defined by
\be
2\mathbf z = \vec \delta + \tau \vec \epsilon\ .
\ee
This modified Theta function  has the following periodicity properties, for $\lambda, \mu\in \Z^g$
\bea
\Theta\le[{\vec \epsilon \atop \vec \delta}\ri] (\mathbf z+\lambda  + \tau\mu)  = \exp\le[ i\pi (\vec \epsilon \cdot \lambda - \vec \delta \cdot \mu) -i\pi \mu\cdot \tau \cdot \mu  - 2i\pi \mathbf z\cdot \mu \ri]
\Theta\le[{\vec \epsilon \atop \vec \delta}\ri] (\mathbf z)\ .
\eea
\bd
The prime form $E(p,q)$ is the $(-1/2,-1/2)$ bi-differential on
$\mathcal L\times \mathcal L$
\bea
E(p,q) &\& = \frac {\Theta_\Delta
  (p-q)}
{h_\Delta (p) h_{\Delta}  (q) }\ ,
\\
&& h_{\Delta} (p)^2 := \sum_{k=1}^{g}
\pa_{\mathfrak u_k}\ln\Theta_\Delta\bigg|_{\mathfrak
  u=0} \omega_k(p) =: \omega_\Delta(p) \label{omegadelta}\ ,
\eea
where $\Delta
= \le[\vec \epsilon\atop \vec \delta \ri]
$ is a half--integer odd characteristic (i.e. $\vec \epsilon\cdot \vec \delta$ is odd). The prime form does
not depend on a choice of $\Delta$.
\ed
The prime form $E(p,q)$ is antisymmetric in the argument and
it is a section of an appropriate line bundle, i.e. it is multiplicatively multivalued on $\mathcal L\times \mathcal L$ :
\bea
&&
E(p + \alpha_j,q) = E(p,q) \ ,\\
&& E(p + \beta_j, q) = E(p,q) \exp{ \le(-\frac {\tau_{jj}}2  - \int_p^{q} \omega_j\ri)}\ .
\eea
In our notation for the half-characteristics, the vectors $\vec \epsilon, \vec \delta$ appearing in the definition of the prime form are actually integer valued.
We also note that the half order differential $h_\Delta$ is in fact also multivalued according to
\bea
h_\Delta(p+\alpha_j) = {\rm e}^{i\pi \epsilon_j} h_\Delta(p)\ ,\\
h_\Delta(p +\beta_j) ={\rm e}^{-i\pi \delta_j} h_\Delta(p).
\label{basespinor}
 \eea 
 Given a meromorphic function $F$ (or a section of a line bundle) we will use the notation $(F)$ for its divisor of zeroes/poles. For example,
\be
(F)\geq -k p + m q\label{divisornotation}
\ee
means that $F$ has at most a pole of order $k$ at $p$ and  a zero of multiplicity at least $m$ at $q$.

\def\cprime{$'$}

%

\end{document}